\documentclass[journal]{IEEEtran}
\IEEEoverridecommandlockouts

\usepackage{graphicx, subcaption}
\usepackage{amsmath,amsthm,amsfonts,amssymb}
\usepackage{cite}
\usepackage{bm}
\usepackage{bbm}
\usepackage{url}
\usepackage{array}
\usepackage{color}
\usepackage{multirow}
\usepackage{booktabs,multirow}
\usepackage{xcolor}
\usepackage{enumerate}
\usepackage{enumitem}
\usepackage{makecell}
\usepackage{comment}

\usepackage[english]{babel}
\usepackage[linesnumbered,vlined,ruled]{algorithm2e}
\usepackage{algpseudocode}

\theoremstyle{plain}

\usepackage{mathtools}

\usepackage[english]{babel}

\usepackage[linesnumbered,vlined,ruled]{algorithm2e}

\allowdisplaybreaks[4]

\DeclareMathOperator*{\argmin}{arg\,min}

\begin{document}
\title{Over-the-Air Learning-based Geometry Point Cloud Transmission}

\author{
Chenghong~Bian,~\IEEEmembership{Student Member,~IEEE},
Yulin~Shao,~\IEEEmembership{Member,~IEEE}, and 
Deniz~G\"und\"uz,~\IEEEmembership{Fellow,~IEEE}
\thanks{C. Bian and D. G\"und\"uz are with the Department of Electrical and Electronic Engineering, Imperial College London, U.K. (e-mails: \{c.bian22,d.gunduz\}@imperial.ac.uk). Y. Shao is with the Department of Electrical and Electronic Engineering, The University of Hong Kong, Hong Kong, China (e-mail: ylshao@hku.hk).
}
\thanks{This work received funding from the UKRI for the projects AI-R (ERC Consolidator Grant, EP/X030806/1) and INFORMED-AI (EP/Y028732/1), and the SNS JU project 6G-GOALS under the EU’s Horizon program (Grant Agreement No. 101139232).}
\thanks{For the purpose of open access, the authors have applied a Creative Commons Attribution (CCBY) license to any Author Accepted Manuscript version arising from this submission.}
\thanks{This paper has been presented in part at the IEEE International Workshop on Signal Processing Advances in Wireless Communications, 2024 \cite{sept}.}

}

\maketitle
\thispagestyle{empty}
\begin{abstract}

3D point cloud is a three-dimensional data format generated by LiDARs and depth sensors, and is being increasingly used in a large variety of applications from autonomous vehicles to robotics and metaverse. This paper presents novel solutions for the efficient and reliable transmission of point clouds over wireless channels {for real-time applications}. We first propose {SEmatic Point cloud Transmission (SEPT)} for small-scale point clouds, which encodes the point cloud via an iterative downsampling and feature extraction process. At the receiver, SEPT decoder reconstructs the point cloud with latent reconstruction and offset-based upsampling. A novel channel-adaptive module is proposed to allow SEPT to operate effectively over a wide range of channel conditions. 
Next, we propose OTA-NeRF, a scheme inspired by neural radiance fields.
OTA-NeRF performs voxelization to the point cloud input and learns to encode the voxelized point cloud into a neural network. Instead of transmitting the extracted feature vectors as in SEPT, it transmits the learned neural network weights over the air in an analog fashion along with few hyperparameters that are transmitted digitally. At the receiver, the OTA-NeRF decoder reconstructs the original point cloud using the received noisy neural network weights. To further increase the bandwidth efficiency of the OTA-NeRF scheme, a fine-tuning algorithm is developed, where only a fraction of the neural network weights are retrained and transmitted. 
{Noticing the poor generality of the OTA-NeRF schemes where the neural network weights are trained for a specific point cloud, we propose an alternative approach, termed OTA-MetaNeRF, which encodes different input point clouds into the latent vectors with shared neural network weights.
Extensive numerical experiments confirm that the proposed SEPT, OTA-NeRF and OTA-MetaNeRF schemes achieve superior or comparable performance over the conventional approaches, where an octree-based or a learning-based point cloud compression scheme is concatenated with a channel code. As an additional advantage, all schemes mitigate the cliff and leveling effects making them particularly attractive for highly mobile scenarios. Finally, the run-time complexities of the schemes are evaluated to verify the capability of the proposed schemes for real-time communications.}
\end{abstract}

\begin{IEEEkeywords}
Deep joint source channel coding (DeepJSCC), point cloud, over-the-air, neural radiance field (NeRF).
\end{IEEEkeywords}

\section{Introduction}\label{sec:intro}
3D point clouds are collections of three-dimensional data points and their associated attributes, such as color, temperature, and normals \cite{Rusu_ICRA2011_PCL,qi2017pointnet,ptv1}.
Generated through technologies such as light detection and ranging (LiDAR), depth camera, and structured light scanning, point clouds are non-ordered and non-uniformly distributed within space. Point clouds have found applications in a wide range of scenarios from environmental monitoring to biomedical imaging, autonomous driving and agriculture.

In many cases, point clouds are captured by remote devices, and are communicated for processing where the necessary computing resources are available. Therefore, wireless transmission plays a vital role in facilitating the mobility and accessibility of 3D point clouds, empowering industries and applications reliant on this expressive data format. 
The standard approach for point cloud transmission consists of two main steps, where octree-based point cloud compression \cite{Rusu_ICRA2011_PCL} is followed by digital transmission over the wireless link.
The standard approach faces several challenges in achieving accurate and reliable transmission of 3D point cloud data, namely, inefficient feature extraction due to the limitation of the octree representation in extracting contextual features, and the cliff and leveling effects, which are inherent to the digital transmission of extracted features.

There has been significant progress in recent years thanks to the adoption of learning-based algorithms and neural network architectures  for point cloud classification, semantic segmentation \cite{qi2017pointnet, thomas2019kpconv, ptv1, guo2021pct, fpt}, and compression \cite{OctSqueeze,pcgcv2, wiesmann2021deep,pct_pcc}. In particular, the authors in \cite{qi2017pointnet} propose PointNet that uses permutation-invariant operations, such as pointwise multilayer perceptrons (MLPs) and max-pooling, to extract features for point cloud classification and segmentation. The follow-up works improve the classification and segmentation performances by using more advanced operations such as 3D convolution (KPconv) \cite{thomas2019kpconv} and self-attention \cite{ptv1,guo2021pct,fpt}. Meanwhile, various deep learning-based schemes have been proposed for point cloud compression \cite{OctSqueeze,pcgcv2, wiesmann2021deep,pct_pcc, ruan2024pointcloudcompressionimplicit, nvfpcc}. Among them, the authors in \cite{OctSqueeze} use neural networks to predict the occupancy probability for a certain node in the octree and adopt an entropy model to generate the compression output. Ref. \cite{wiesmann2021deep} uses multiple KPconv \cite{thomas2019kpconv} and downsampling layers to progressively extract information, and adopts an offset-based up-sampling algorithm to reconstruct the point cloud at the decoder. 
The authors in \cite{pct_pcc} follow the framework in \cite{wiesmann2021deep} and use the point cloud transformer \cite{guo2021pct} as the backbone to enhance the compression performance. 

Recently, neural radiance fields (NeRFs) \cite{NeRF, coin} have presented a new perspective for point cloud compression.
Originally proposed in \cite{NeRF}, NeRF aims to represent a 3D scene using a neural network whose weights are trained to overfit a single input sample. Once trained, the NeRF model is capable to provide high quality reconstructions of the same source from different angles. 
Following the idea of NeRF,  the authors in \cite{nvfpcc} represent each part of the point cloud using a neural network along with a latent code. The neural network is trained to predict the occupancy information of a voxel given its x, y and z coordinates. The follow-up work \cite{ruan2024pointcloudcompressionimplicit} improves the rate-distortion (R-D) performance in \cite{nvfpcc} by adopting the state-of-the-art neural network compression algorithm \cite{deepcabac} and extends the framework to tackle more challenging attribute compression of the point cloud. Both schemes are reported to outperform the benchmark geometry-based point cloud compression (G-PCC) algorithm \cite{GPCC}. {Noticing the poor generalization of prior schemes in \cite{coin, nvfpcc, ruan2024pointcloudcompressionimplicit}, instead of transmitting the compressed neural network weights, the authors in \cite{coinpp} propose transmitting the modulations applied to a meta-learned base network as its compression output. Once deployed, different modulations are generated for different data samples achieving superior reconstruction performance and significantly higher encoding efficiency.}

Another direction to facilitate the wireless transmission of point clouds is to employ joint source-channel coding \cite{gündüz2024jointsourcechannelcodingfundamentals, JSCC2019}, utilizing analog transmission to mitigate the cliff and leveling effects, which the conventional separation-based architectures suffer from. Over the years, there has been a growing interest in utilizing DeepJSCC to develop semantic communication systems \cite{Deniz2022}. 
In particular, researchers have actively applied DeepJSCC to different wireless channels, e.g., multi-path fading \cite{jsccofdm}, MIMO \cite{vit_mimo}, and relay channels \cite{pf_relay_jscc}, as well as different data modalities, e.g., image \cite{bupt}, video \cite{deepwive} and speech \cite{deepjscc_speech}. Recently, it is shown in \cite{airnet} that DeepJSCC is also applicable to efficient transmission of neural network weights over the wireless channel. In particular, the authors in \cite{airnet} train a neural network, dubbed AirNet, which takes both the classification accuracy as well as the noisy wireless channel into account and formulate it as a DeepJSCC problem. The neural network weights are directly transmitted over the channel thus avoiding the cliff and leveling effects. It is shown that AirNet achieves higher classification accuracy compared with the digital baselines using less number of channel uses.

Despite the importance of wireless delivery in enabling many point cloud applications, only a few papers investigate the wireless point cloud transmission problem \cite{sept, pointcloudvideo, HoloCast+, gnn_pct, jap_inr}. 
In particular, our previous work, \cite{sept} directly maps the point cloud into a latent vector, which is transmitted directly over the channel. The authors in \cite{HoloCast+} adopts a hybrid digital-analog transmission approach, where the point cloud is first compressed using octree-based point cloud compression algorithm and the compressed bits are transmitted using digital coded modulation. They then adopt the graph Fourier transform to encode the residual information, which is transmitted in an analog fashion. The same authors further introduce a graph convolutional neural network for more efficient geometric feature extraction in \cite{gnn_pct}, and  employ the NeRF method to efficiently encode the graph information in \cite{jap_inr}. Though digital coded modulation is still required to deliver side information for successful reconstruction at the decoder, these schemes can achieve modest reconstruction performance while mitigating cliff and leveling effects.  {Another related work is \cite{ext_sept}, where the authors extend our previous work \cite{sept} by digitally transmitting additional coordinate information to achieve superior reconstruction performance.} 

\begin{figure*}[t]
  \centering
  \includegraphics[width=0.9\linewidth]{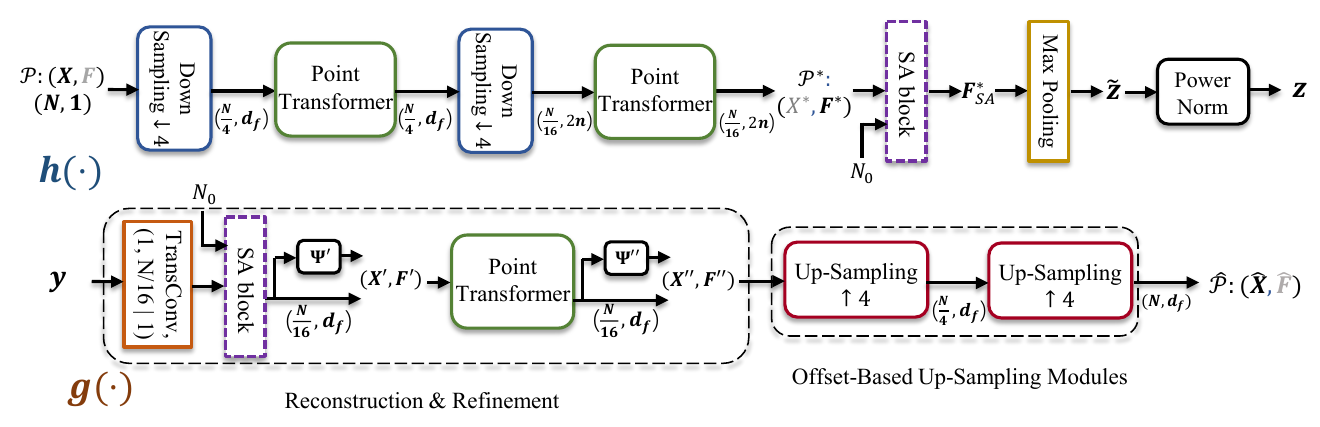}\\
  \caption{The detailed architecture of SEPT with an encoder $h(\cdot)$ and a decoder $g(\cdot)$. For clarity, we label the dimension information of the point cloud after each processing module. As an example, in $(\frac{N}{4},d_f)$ after the first point transformer at the encoder, $\frac{N}{4}$ refers to the number of points after the point transformer and $d_f$ is the dimension of the feature vectors. The coordinate reconstruction modules as denoted as $\bm{\Psi}^\prime$ and $\bm{\Psi}^{\prime \prime}$. {The SA block, which takes the feature tensor and the noise variance $N_0$ as input, is designed for SNR-adaptive point cloud transmission.}}
\label{fig:system_model}
\end{figure*}

In this paper, we first leverage the recent advances in point cloud processing \cite{ptv1, thomas2019kpconv}, especially in point cloud compression \cite{pcgcv2, pct_pcc, wiesmann2021deep}, in the context of deep joint source-channel coding (DeepJSCC) \cite{JSCC2019} to develop the SEPT (SEmantic Point cloud Transmission) framework, which is designed for small-scale point cloud transmission over bandwidth limited channels. 
At the transmitter, SEPT encoder maps the point cloud directly into a latent vector without voxelization using the point transformer \cite{ptv1} as its backbone. At the receiver, instead of feeding the noisy latent vector directly into the up-sampling layers, we introduce a refinement layer to first denoise the latent vector. Finally, offset-based up-sampling layers are employed for point cloud reconstruction. We further introduce a signal-to-noise-ratio (SNR)-adaptive (SA) module for adaptation to channel variations, through which we train a single pair of SEPT encoder and decoder that can operate over a wide range of channel conditions. {Given that points within a point cloud are distributed irregularly in the 3D space, which significantly differs from images, where pixels lie on a regular 2D grid. Therefore, designing the SA and refinement modules of the SEPT model is non-trivial and is essentially different from previous DeepJSCC schemes tailored for image transmission.}

{While SEPT is designed mainly for small scale point clouds, we further propose the OTA-NeRF and OTA-MetaNeRF frameworks for the transmission of more complex point clouds. Both schemes first voxelize the input point cloud and feed the coordinates of the voxels to the corresponding neural networks to produce the occupancy probabilities for reconstruction. For the OTA-NeRF scheme, a specific neural network is trained to represent each point cloud input, whose weights are transmitted directly over the wireless channel. Similarly to AirNet \cite{airnet}, noise injection during training is used to enable robustness against channel noise. 
Note that, in AirNet, the model is trained on datasets, and the objective is to obtain models that can generalize to new samples which clearly distinguish from our OTA-NeRF scheme. To further increase the bandwidth and encoding efficiency, a fine-tuned OTA-NeRF scheme is proposed, where we only finetune and transmit the difference in the neural network weights with respect to a base model that is available to the receiver. The OTA-MetaNeRF model, on the other hand, adopts a shared pretrained architecture, comprised of SIREN layers \cite{coinpp} and ModulationNet. Latent vectors corresponding to different data samples are transmitted over the noisy channel and fed to the ModulationNet, whose outputs are adopted to modulate the hidden features produced by the SIREN layers. Latent vectors are optimized via gradient descent, which converges within a few steps.  }

Our main contributions can be summarized as follows:
\begin{itemize}[leftmargin=0.4cm]
    \item We present SEPT for the efficient delivery of small-scale 3D point clouds over static and fading wireless channels.  The SEPT encoder directly maps the input point cloud into a latent vector with fully analog transmission. We show that SEPT achieves significant performance gains over the digital benchmarks for the delivery of small-scale point clouds over bandwidth-limited wireless links, targeting real-time applications.
    \item We introduce novel SNR-adaptive encoding and decoding modules for SEPT. This module differs from the attention-based designs in \cite{xu2021wireless, vit_mimo} for SNR-adaptive image transmission. In particular, the mean-pooling operation calculates the average of the features belonging to different points following the permutation invariance of point clouds.
    \item To facilitate the transmission of larger point clouds, the OTA-NeRF framework is proposed, where each point cloud input is represented by a specific neural network. The neural network is trained to minimize the reconstruction loss, and its weights are transmitted over the wireless channel in an analog fashion along with a limited number of hyperparameters,  which are transmitted digitally using channel codes.
    \item To further improve the performance of the OTA-NeRF framework, we consider a fine-tuning approach, where, to transmit different point clouds, only a fraction of the neural network weights need to be re-trained and transmitted, leading to an improved bandwidth-distortion trade-off, as well as a reduction in the encoding complexity.
    \item {A novel meta-learning based OTA-MetaNeRF scheme is proposed to simultaneously improve the reconstruction performance, the encoding efficiency and the generalization capability of OTA-NeRF. OTA-MetaNeRF encodes different point clouds by modulating a single common neural network whose parameters are fixed and shared with the receiver. Instead, different samples are represented using different latent vectors, which are obtained via gradient backpropagation, and transmitted over the channel. }
    \item {Extensive simulations verify that the proposed SEPT, OTA-NeRF and OTA-MetaNeRF schemes outperform or achieve comparable reconstruction performance and computation latency with respect to (w.r.t.) the fully digital benchmarks that first compress the point cloud using G-PCC or other learning-based algorithms, and transmit the compressed bits using digital coded modulations. These confirm the capability of the proposed schemes for real-time point cloud transmission. More importantly, all schemes  eliminate/mitigate the cliff and leveling effects, which makes them particularly appealing to be deployed in highly mobile scenarios, where accurate channel estimation is not feasible or very costly. }
\end{itemize}

\section{System Model}\label{sec:II}

A point cloud can be expressed as $\mathcal{P} = (\bm{X}, \bm{F})$, where $\bm{X}\triangleq \{\bm{x}_i\in \mathbb{R}^3\}$, $i\in [N]$, is a set of $N$ points in space, and $\bm{F}\triangleq \{\bm{f}_i\in \mathbb{R}^{d}\}, i\in [N]$, is a set of features associated with each point in $\bm{X}$. 
In this paper, we consider point clouds with no input attributes\footnote{Nevertheless, the point clouds in the intermediate layers of SEPT can have non-trivial attributes/features. For example, the neighboring information is contained in the attributes of the downsampled points.} and focuses on transmitting only the coordinates in $\bm{X}$.
Following the convention, we set the input attributes $\bm{F}$ to an all-ones vector with $d = 1$.

The encoder maps the input 3D point cloud $\mathcal{P}$ to a channel codeword, $\bm{{z}} \in \mathbb{C}^n$, where $n$ is the available channel bandwidth. The channel use per point (CPP) is given by $\frac{n}{N}$. We impose a power constraint on transmitted codewords:
\begin{equation}
    {\frac{1}{n}\|\bm{{z}}\|^2_2 \le P},
    \label{equ:def_z}
\end{equation}
{where $P$ denotes the average power per complex symbol.}
The received signal $\bm{{y}}^\prime\in\mathbb{C}^{n}$ is expressed as:
\begin{align}
    \bm{{y}}^\prime = h \bm{{z}} + \bm{v},
    \label{eq:awgn_channel}
\end{align}
where $\bm{v}\in\mathbb{C}^{n}$ denotes a complex additive white Gaussian noise (AWGN) vector with independent and identically distributed elements, {$\bm{v}\sim\mathcal{CN}(\bm{0},N_0 \bm{I}_{n})$}. When $h$ is a constant, the wireless channel in \eqref{eq:awgn_channel} degrades to an AWGN channel.  The average SNR is defined as $\mathrm{SNR} = 10\log_{10}{\frac{|h|^2P}{N_0}}$.  {Without loss of generality}, we assume $P=1$ in the sequel.

We also consider a slow fading channel, where $h$ is modeled as a random variable, which takes an independent realization from an underlying distribution for the transmission of each point cloud. We assume that both the transmitter and the receiver have perfect knowledge on the channel state information (CSI), $h$.
Upon receiving $\bm{{y}}^\prime$, the decoder equalizes it using $h$ as in \cite{pf_relay_jscc}. Then, the equalized signal is converted to a real vector $\bm{y}\in\mathbb{R}^{2n}$ which is utilized in the subsequent decoding process to obtain the reconstructed point cloud $\hat{\mathcal{P}}$. 

Two conventional peak signal-to-noise ratio (PSNR) measures \cite{D1D2}, $\text{D1}$ and $\text{D2}$, are adopted to evaluate the reconstruction quality. For point clouds $\mathcal{A}$ and $\mathcal{B}$, $\text{D1}(\mathcal{A}, \mathcal{B})$ measures the average point-to-point geometric distance {{between each point in point cloud $\mathcal{A}$ and its nearest neighbor in point cloud $\mathcal{B}$. To be precise, we first calculate the mean squared error, $e_{\mathcal{A},\mathcal{B}}^{D1}$:
\begin{align}
    e_{\mathcal{A},\mathcal{B}}^{D1} & \triangleq \frac{1}{N} \sum_{\bm{a}_i \in \mathcal{A}}||\bm{a}_i-\bm{b}_{k}||_2^2, \notag \\ 
   \text{where} \quad k &= \argmin_{j\in [N]}{||\bm{a}_i - \bm{b}_j||_2^2}.
    \label{eq:MSE_D1}
\end{align}
Then, $\text{D1}$ is defined as \cite{D1D2}:
\begin{equation}
    \text{D1}(\mathcal{A}, \mathcal{B}) \triangleq 10 \frac{3 \gamma^2}{\max \left (e_{\mathcal{A},\mathcal{B}}^{D1}, e_{\mathcal{B},\mathcal{A}}^{D1} \right)},
    \label{eq:D1}
\end{equation}
where factor $3$ in the numerator is due to the 3D coordinates used in the representation, and the peak value, $\gamma$, is set to unity due to the fact that the input points are normalized within the range $[0, 1]$.
Similarly, $\text{D2}$ evaluates the point-to-plane distance between point clouds $\mathcal{A}$ and $\mathcal{B}$ and the error term for $\text{D2}$ is defined as:
\begin{equation}
    e_{\mathcal{A},\mathcal{B}}^{D2} = \frac{1}{N} \sum_{\bm{a}_i \in \mathcal{A}}(\bm{a}_i-\bm{b}_k)\cdot\bm{n}_i ,
    \label{eq:D2}
\end{equation}
where $\bm{n}_i$ is the normal vector corresponding to $\bm{a}_i$ and $\bm{b}_k$ is the nearest neighbor of $\bm{a}_i$ in $\mathcal{B}$. After obtaining \eqref{eq:D2}, we follow the same formula in \eqref{eq:D1} to calculate $\text{D2}(\mathcal{A}, \mathcal{B})$.}}

\section{SEPT framework}\label{sec:III}

\begin{figure*}
     \centering
     \begin{subfigure}{0.4\columnwidth}
         \centering
         \includegraphics[width=0.9\columnwidth, height=1.13\columnwidth]{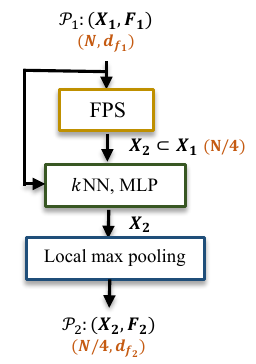}
         \caption{}
     \end{subfigure}
     \hspace{0.2cm}
     \begin{subfigure}{0.56\columnwidth}
         \centering\includegraphics[width=\columnwidth,height=0.82\columnwidth]{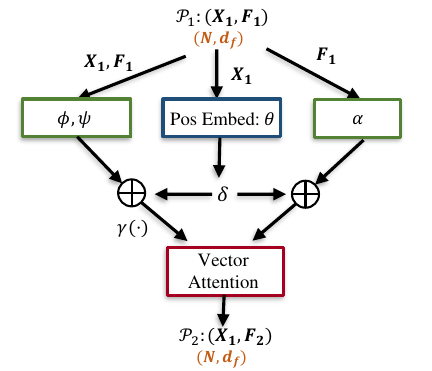}
         \caption{}
     \end{subfigure}
      \hspace{0.2cm}
     \begin{subfigure}{0.6\columnwidth}
         \centering
         \includegraphics[width=\columnwidth, height=0.67\columnwidth]{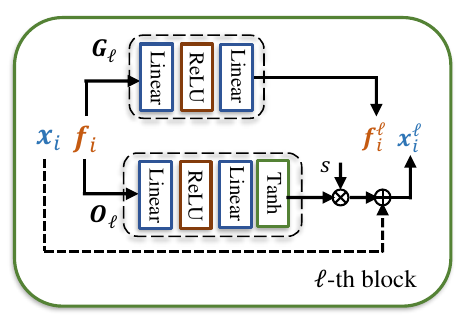}
         \caption{}
     \end{subfigure}
  \caption{The inner architectures of (a) the downsampling module,  (b) the point transformer layer, and (c) the $\ell$-th block of the offset-based up-sampling module, $\ell\in [1,L]$.}
\label{fig:detailed_net}
\end{figure*}

This section details the design of the {encoding and decoding} functions for the SEPT framework, originally proposed in \cite{sept}. 
{To make the SEPT framework practically relevant, we also introduce a novel SNR-adaptive module for SEPT to achieve the reconstruction performance obtained by separately trained models.}  The detailed architecture of SEPT is presented in Fig.~\ref{fig:system_model}.

\subsection{SEPT Encoder}\label{sec:IIIA}
The encoder of SEPT consists of three main modules: downsampling, self-attention, and max pooling.

\textbf{Downsampling.} The objective of the downsampling module is to reduce the number of points in the input point cloud.
Let $\mathcal{P}_1=(\bm{X}_1, \bm{F}_1)$ and $\mathcal{P}_2=(\bm{X}_2, \bm{F}_2)$ denote the input and output point clouds of a downsampling module, respectively, where $\bm{X}_2 \subset\bm{X}_1$. 
As shown in Fig. \ref{fig:detailed_net}(a), the proposed SEPT uses the farthest point sampling\footnote{{In principle, we can directly generate the encoded vectors without using the FPS. However, this would cause significant computation complexity and performance loss which is observed in our experiments.}} (FPS) algorithm to generate the down-sampled geometric information, $\bm{X}_2$. To generate the feature vector of the $i$-th point in $\bm{X}_2$, denoted by $\bm{f}^{(2)}_i\in \mathbb{R}^{d_{f_2}}$, we first find its $k$-nearest neighbors within a given radius $r$ in $\bm{X}_1$, denoted by $\mathbb{N}(i)$.\footnote{If the $i$-th ($i<k$) neighbor has a distance larger than $r$ from the sampled point, we will use the nearest neighbor to replace it.}
{Then, we concatenate the feature vectors of the points in $\mathbb{N}(i)$ with their coordinates and organize them into a tensor, {denoted by $\bm{T}_{\mathbb{N}(i)} \in \mathbb{R}^{(d_{f_1}+3)\times k \times N_2}$, where $N_2$ denotes the cardinality of $\bm{X}_2$,}} and feed this tensor to a 2D convolutional layer followed by batch normalization, ReLU, and max pooling. In each downsampling module, we set the cardinality of $\bm{X}_2$ to be $1/4$ of that of $\bm{X}_1$, i.e., {$|\bm{X}_2| = |\bm{X}_1|/4$}.

\textbf{Self-Attention Module.} 
In SEPT, each downsampling module is followed by a self-attention module \cite{ptv1}, a point cloud processing technique that is capable of extracting rich neighboring information. 
Denote the input and output point clouds of the self-attention layer by $\mathcal{P}_1=(\bm{X}_1, \bm{F}_1)$ and $\mathcal{P}_2=(\bm{X}_1, \bm{F}_2)$, respectively. As shown in Fig. \ref{fig:detailed_net}(b), the self-attention layer refines the features of each point in $\mathcal{P}_1$. The inner operations can be written as
\begin{align}
    \bm{f}^{(2)}_i = \sum_{j\in \mathbb{N}(i)} \!\text{Softmax}&\big(\gamma(\phi(\bm{f}^{(1)}_i) - \phi(\bm{f}^{(1)}_j))+\delta(\bm{x}_i, \bm{x}_j)\big) \notag \\ \odot \; &(\alpha(\bm{f}^{(1)}_j) + \delta(\bm{x}_i, \bm{x}_j)),
\end{align}
where $\bm{f}^{(1)}_i$ and $\bm{f}^{(2)}_i$ denotes the original and the refined  feature vectors of the $i$-th point, respectively;
$\gamma, \phi, \alpha: \mathbb{R}^{d_f} \rightarrow \mathbb{R}^{d_f}$ are realized by MLPs where $d_f$ is the feature dimension; 
$\delta(\bm{x}_i, \bm{x}_j) = \theta(\bm{x}_i-\bm{x}_j)$ is the positional information and $\theta: \mathbb{R}^{3} \rightarrow \mathbb{R}^{d_f}$ is an MLP layer for positional embedding;
$\odot$ denotes the element-wise product. That is, we adopt vector attention \cite{ptv1}, as opposed to the standard scalar dot-product attention used in language and vision transformer models, {for better performance}.

As shown in Fig. \ref{fig:system_model}, after two consecutive downsampling and self-attention module pairs, the final downsized point cloud is obtained, which we denote by $\mathcal{P}^* = (\bm{X}^*,\bm{F}^*)$, where $\bm{X}^*\in\mathbb{R}^{(N/16)\times 3}$ and $\bm{F}^*\in\mathbb{R}^{(N/16)\times 2n}$. Note that $n$ is the length of the transmitted codeword, $\bm{z}$ in \eqref{equ:def_z}.
We emphasize that the features $\bm{F}^*$ are generated by neural networks and can be optimized to be robust to noise, thanks to end-to-end learning. 
On the other hand, the coordinates $\bm{X}^*$ are obtained from FPS and are susceptible to noise.
Our empirical results indicate that $\bm{X}^*$ has to be transmitted to the receiver reliably via digital communications. 
Failure to do so results in a substantial degradation in the reconstruction performance of the point cloud. 
Digital transmission of $\bm{X}^*$, however, would cause two problems:
1) the cliff and leveling effects; and
2) excessive channel usage.
To mitigate these problems, SEPT completely eliminates the need to transmit the coordinate tensor {($\bm{X}^*$)}, and instead focuses solely on transmitting the features {($\bm{F}^*$)} to the receiver. 
To be precise, SEPT learns to encode the global features in $\bm{F}^*$ as the transmitted channel codeword, and the decoder is trained to reconstruct the entire point cloud from the noisy channel output without the aid of its coordinate tensor.
By doing so, SEPT significantly reduces the amount of data that needs to be transmitted, leading to more efficient and streamlined communication.

\textbf{Max Pooling.} 
The last step\footnote{The input to the max pooling module will be $\bm{F}^*_{SA}$ if the SA module is adopted.} at the encoder is to transform $\bm{F}^*\in \mathbb{R}^{(N/16)\times 2n}$ to the latent vector $\tilde{\bm{z}}\in\mathbb{R}^{2n}$.
This is done by applying max pooling over the $N/16$ points to generate the $2n$-dimensional vector $\tilde{\bm{z}}$ which will be further converted to the complex codeword $\bm{z} \in \mathbb{C}^n$ defined in \eqref{equ:def_z}.

\subsection{SEPT Decoder}\label{sec:IIIB}
The decoder of SEPT consists of two modules: latent reconstruction and refinement, and offset-based up-sampling. 

\textbf{Latent Reconstruction and Refinement.} 
The latent reconstruction module takes the noisy channel output as input to reconstruct $\mathcal{P}^*$.
As shown in Fig. \ref{fig:system_model}, given the received signal $\bm{y}$, we first use a TransConv layer, which is essentially a 1D deconvolution with a unit stride, to generate the initial estimate of $\bm{F}^*$, denoted by  ${\bm{F}}^{\prime} \in \mathbb{R}^{(N/16) \times 2n}$.
Then, we employ a coordinate reconstruction module $\Psi^\prime: \mathbb{R}^{2n} \rightarrow \mathbb{R}^{3}$, which is composed of MLP layers and a ReLU function, {and operates on each row of ${\bm{F}}^{\prime}$} to generate an initial estimate of the coordinates:
\begin{align}
    {\bm{X}}^{\prime} = \Psi^\prime({\bm{F}}^{\prime}),
    \label{eq:cood_recon}
\end{align}
where ${\bm{X}}^{\prime}\in \mathbb{R}^{(N/16) \times 3}$.
The initial estimates $({\bm{X}}^{\prime},{\bm{F}}^{\prime})$ can be erroneous due to noise. Therefore, we further use a self-attention module, to refine the features:
\begin{align}
    {\bm{F}}^{\prime\prime} = \text{Self-Attention}({\bm{X}}^{\prime}, {\bm{F}}^{\prime}).
    \label{eq:refine}
\end{align}
Next, a new coordinate reconstruction layer, $\Psi^{\prime\prime}$, is applied to ${\bm{F}}^{\prime\prime}$ to produce a refined estimation of coordinates, ${\bm{X}}^{\prime\prime}$.
Our refinement module is shown to be very effective in denoising the coordinates and features.

\textbf{Offset-Based Up-Sampling.} Finally, we employ an  offset-based up-sampling module \cite{wiesmann2021deep} on $({\bm{X}}^{\prime\prime},{\bm{F}}^{\prime\prime})$ for point cloud reconstruction. For the $i$-th point in the input point cloud, whose coordinates and features are denoted by $(\bm{x}_i, \bm{f}_i)$, this module generates $L$ new points $\{(\bm{x}_i^\ell, \bm{f}_i^\ell), \ell\in [1, L]\}$ as:
\begin{align}
    \bm{x}_i^\ell &= \bm{x}_i + s \cdot O_\ell(\bm{f}_i), \\
    \bm{f}_i^l &= G_\ell(\bm{f}_i),
    \label{eq:upsample}
\end{align}
where $O_\ell: \mathbb{R}^{d_f} \rightarrow [-1, 1]^3$ is an MLP layer followed by a $\mathrm{tanh}$ function that aims to generate an offset vector;
$G_\ell$ is comprised of MLPs and a ReLU function that maps the input feature to a new one with the same dimension; and $s$ is a scaling factor for the offsets. 
Compared with \cite{wiesmann2021deep}, SEPT uses a relatively large scaling factor $s = 0.1$ to give the up-sampling module more freedom for better performance, considering the additional noise introduced by the wireless channel.
The detailed architectures for $O_\ell$ and $G_\ell$ are shown in Fig. \ref{fig:detailed_net}(c).

In SEPT, we use two up-sampling modules, and $L$ is set to $4$ in each module to match the encoder operations. 
Denoting by $\hat{\mathcal{P}} =  (\hat{\bm{X}}, \hat{\bm{F}})$ the final output of the up-sampling modules, the Chamfer distance between ${\bm{X}}$ and $\hat{\bm{X}}$, denoted by $d^2_{cd}({\bm{X}}, \hat{\bm{X}})$, is used as the loss function:
\begin{align}
    { \mathcal{L}_{\text{CD}} = \frac{1}{N} \sum_{\bm{x}\in \mathcal{P}} \min_{\hat{\bm{x}} \in \hat{\mathcal{P}}} ||\bm{x} - \hat{\bm{x}}||^2_2 + \frac{1}{N} \sum_{\hat{\bm{x}}\in \hat{\mathcal{P}}} \min_{\bm{x}\in {\mathcal{P}}} ||\hat{\bm{x}} - \bm{x}||^2_2.}
    \label{eq:chamfer_dist}
\end{align}

\begin{figure}[t]
\centering
\includegraphics[width=0.9\linewidth]{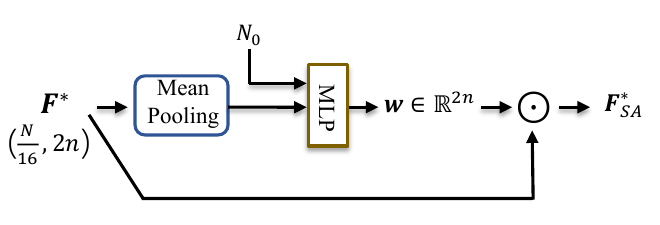}
\caption{{The proposed SA module for SNR-adaptive point cloud transmission.}}
\label{fig:SA_module}
\end{figure}

\subsection{SNR-adaptive (SA) Module}
{We next describe the design for our SNR-adaptive module, which is shown in Fig.\ref{fig:system_model}. 

At the SEPT encoder, the SA block takes the feature tensor, $\bm{F}^* \in \mathbb{R}^{\frac{N}{16}\times 2n}$ as input and adopts mean pooling along the first dimension. This generates an $n$-dimensional vector which is concatenated with the noise variance, $N_0$, and fed to an MLP layer, whose output is a weight vector $\bm{w} \in \mathbb{R}^{2n}$. The vector is repeated $\frac{N}{16}$ times along the first {dimension} to obtain a matrix $\bm{W} \in \mathbb{R}^{\frac{N}{16}\times 2n}$, which will be multiplied with $\bm{F}^*$ in an element-wise manner. {The entire operation at the SEPT encoder is shown in Fig. \ref{fig:SA_module} and summarized as follows:}
\begin{align}
     \bm{w} = &\text{MLP}\left( \left[\frac{16}{N}\sum_{i=1}^{N/16}\bm{F}^*[i,:], N_0 \right] \right) \notag \\
    &\bm{F}^*_{SA} ={\underbrace{[\bm{w}, \cdots, \bm{w}]}_{N/16}}^T \odot \bm{F}^*.
    \label{equ:SA_block}
\end{align}
We further note that one can also perform the mean pooling operation along the second axis; however, simulation results show that the presented scheme is strictly better. This aligns with the analysis in \cite{qi2017pointnet}; since the points within a point cloud are unordered, calculating the average over $\frac{N}{16}$ points means that the result would not be effected much from the input order of the points in the point cloud. The SA block at the SEPT decoder follows exactly the same processing, and is omitted here due to page limit.
}

\section{OTA-NeRF framework}\label{sec:IV}
The SEPT model presented in the previous section is designed to transmit small point clouds. However, its performance is limited when the input point cloud has a large number of points. This is due to the fact that the max-pooling operation at the SEPT encoder causes substantial loss of details. In this section, we introduce a hybrid point cloud transmission scheme, dubbed OTA-NeRF, which is capable to transmit point clouds of arbitrary size.

\subsection{Representing Point Clouds via Neural Networks}
In this subsection, we first introduce the underlying principle of the OTA-NeRF framework that the data samples such as images and point clouds, can be represented using neural networks \cite{NeRF, ruan2024pointcloudcompressionimplicit}. We start with the training and testing procedures of the OTA-NeRF without considering the noise introduced by the wireless channel.

\subsubsection{Point cloud preprocessing}
As before, we focus on the transmission of geometric information, thus, we consider a point cloud $\mathcal{P} = \{\bm{X}, \bm{F}\}$, where $\bm{F}$ is set to an all-ones vector. {$\mathcal{P}$ is first normalized within the range of $[0, 1]$, i.e., $\bm{x}_i \in [0, 1]^3, \; \forall i \in [1, N]$, where $N$ is the number of points in $\mathcal{P}$}. 

\textbf{Voxelization:}
Unlike images, where pixels lie on a regular grid, the points of a point cloud are distributed irregularly in the 3D space.
To be consistent with the original NeRF framework, we partition the input point cloud $\mathcal{P}$ into voxels. A uniform voxelization is adopted, where each unit interval along the x, y and z axes is equally partitioned into $2^V$ segments. As a result, $2^{3V}$ number of voxels are obtained and each voxel is represented by its indices, $(i, j, k); i, j, k \in [1, 2^V]$ corresponding to the $(i,j,k)$-th segment along the x, y and z axes, respectively. 

Similar to the Octree representation \cite{GPCC}, a voxel is occupied if there exists at least one point in it. By traversing over all the voxels, the original point cloud, $\mathcal{P}$, is transformed into a discrete representation, denoted as $\mathcal{G}^\prime \triangleq \{\mathcal{V}^\prime, \mathcal{O}^\prime\}$, where $\mathcal{V}^\prime$ denotes all the $2^{3V}$ voxels with indices ranging from $(1, 1, 1)$ to $(2^V, 2^V, 2^V)$. The occupancy information of $\mathcal{V}^\prime$ is represented by $\mathcal{O}^\prime$, whose $(i,j,k)$-th element, $\mathcal{O}^\prime_{i,j,k} \in \{0, 1\}$, indicates whether the voxel with indices, $(i, j, k)$, is occupied or not. Finally, we denote the number of occupied voxels as $N_o$.

\textbf{Block Partition:}
In general, the discrete representation, $\mathcal{G}^\prime$ is sparse, i.e., $N_o \ll 2^{3V}$, which motivates us to remove the regions which are empty. 
To be precise, we partition the voxels $\mathcal{V}^\prime$ uniformly into $2^{3B}, (B < V)$ blocks where each block contains $2^{3(V-B)}$ voxels. If none of the voxels within the block are occupied, then we mark it as an empty block. The voxels belonging to the empty blocks are pruned from $\mathcal{V}^\prime$ and  the voxels of the remaining occupied blocks are denoted by $\mathcal{V}$ with cardinality $N_V \triangleq N_B 2^{3(V-B)}$, where $N_B \in [1, 2^{3B}]$ is the number of occupied blocks. We further denote the block occupancy information by $\mathcal{O}_B$. Intuitively, the block partition is essential in improving the reconstruction efficiency as the occupancy of the voxels belonging to the empty blocks are no longer needed to be predicted. After preprocessing, the original point cloud is represented by:
\begin{equation}
    \mathcal{G} \triangleq \{\mathcal{V}, \mathcal{O}, \mathcal{O}_B\},
    \label{eq:discrete_pc}
\end{equation}
where $\mathcal{O}$ represents the occupancy of the voxels within $\mathcal{V}$.

\begin{figure}[t]
     \centering
     \begin{subfigure}{0.9\columnwidth}
         \centering
         \includegraphics[width=\columnwidth]{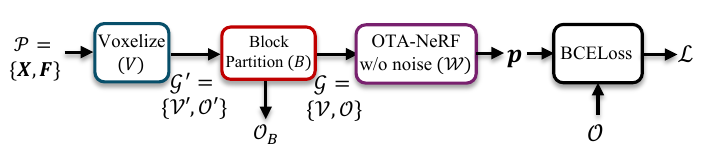}
         \caption{Training phase.}
     \end{subfigure}     
     \begin{subfigure}{0.9\columnwidth}
         \centering
         \includegraphics[width=\columnwidth]{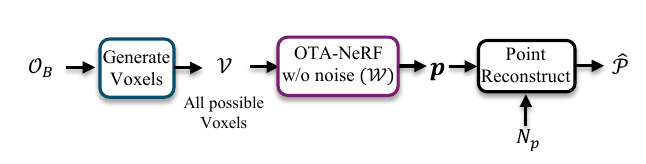}
         \caption{Testing Phase.}
     \end{subfigure}

  \caption{Illustrations of the training and testing phases of the proposed OTA-NeRF framework without noise.}
  \label{fig:point_nerf_system}
\end{figure}

\begin{figure*}[t]
  \centering
  \includegraphics[width=0.8\linewidth]{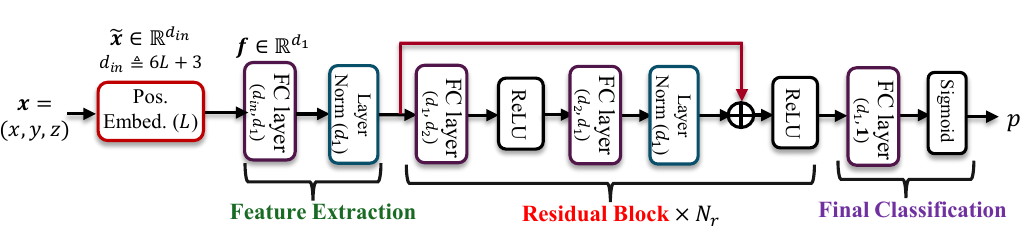}\\
  \caption{The neural network architecture of the OTA-NeRF model. In particular, the model is comprised of four modules, the positional embedding module, the feature extraction module, the residual blocks and the final classification module. The OTA-NeRF model takes the coordinate of a voxel as input and predicts if the voxel is occupied or not.}
\label{fig:point_nerf_model}
\end{figure*}

\subsubsection{Training the neural network}
Next, we train a neural network to represent the point cloud whose detailed architecture is shown in Fig. \ref{fig:point_nerf_model}. In particular, the input to the neural network is a three-dimensional vector, $\bm{x} = (i, j, k)/2^V$, representing the x, y and z coordinates of the voxels belonging to $\mathcal{V}$. Instead of directly feeding $\bm{x}$ to the neural network, we first employ a positional embedding module proposed in \cite{NeRF} to transform it to a higher dimensional vector, denoted by $\tilde{\bm{x}}$:
\begin{align}
    \tilde{\bm{x}} = [\bm{x}, &\sin(2^0 \pi \bm{x}), \cos(2^0 \pi \bm{x}), \ldots, \notag \\
    &\sin(2^{L-1} \pi \bm{x}), \cos(2^{L-1} \pi \bm{x})],
    \label{eq:pos_emb}
\end{align}
where $L$ is the number of different frequencies and the dimension of $\tilde{\bm{x}}$ is denoted by $d_{in} \triangleq 6L + 3$. Note that employing the positional embedding module provides a much richer representation to the subsequent layers compared with the original input $\bm{x}$, yielding a better performance. 

The neural network shown in Fig. \ref{fig:point_nerf_model} is comprised of three learnable modules, namely, the feature extraction, residual blocks, and the final classification module. All these modules employ fully connected (FC) layers, layer normalization (LN) and non-linear activation functions. The feature extraction module transforms the input $\tilde{\bm{x}}$ to a feature vector, $\bm{f} \in \mathbb{R}^{d_1}$. $N_r$ residual blocks with residual connections are employed to improve the classification accuracy. The final classification module is comprised of a FC layer followed by a sigmoid activation function to produce the corresponding probability value, $p$. The neural network is trained to minimize the BCE loss\footnote{We also tried the focal loss as the loss function \cite{ruan2024pointcloudcompressionimplicit}; however, BCE loss yields a better performance in our simulations.} with a goal to correctly predict the occupancy of the voxels:
\begin{align}
    \mathcal{L} = \frac{1}{N_{V}} \sum_{v = 1}^{N_{V}} - (o_v \log(p_v) + (1-o_v) \log(1-p_v)),
    \label{eq:bce_loss}
\end{align}
where $N_V$ is the total number of voxels in $\mathcal{V}$ and $o_v \in \{0, 1\}$ indicates if the $v$-th voxel is occupied or not.

\subsubsection{Point cloud reconstruction}
After training, we assume that the neural network weights, denoted by $\mathcal{W}$, and the block occupancy information, $\mathcal{O}_B$, are available at the decoder. The voxels, $\mathcal{V}$, are first regenerated using the block occupancy information, $\mathcal{O}_B$. Then, the coordinates of the voxels within $\mathcal{V}$ are fed to  the neural network with weights, $\mathcal{W}$, to obtain the probability vector, $\bm{p} \in (0, 1)^{N_V}$.
We then sort the elements of $\bm{p}$ and the voxels corresponding to the top-$N_{p}$ probability values will be marked as occupied while the others are assumed to be empty. Note that $N_{p}$ does not need to be the same with the number of occupied voxels in $\mathcal{G}$, i.e., $N_o$. As will be shown in the simulation part, choosing a proper $N_{p}$ value is essential for a good reconstruction performance.  In real implementations, the encoder generates different reconstructed point clouds, $\hat{\mathcal{P}}$, with different $N_p$ values. The $D1$ performances of these reconstructed point clouds can be calculated by the encoder, and the $N_p$ which yields the best $D1$ performance can be adopted and transmitted to the decoder as a hyper parameter.
Finally, the overall training and testing processes for the OTA-NeRF model without noise are illustrated in Fig. \ref{fig:point_nerf_system}.

\begin{figure}[t]
     \centering
     \begin{subfigure}{0.8\columnwidth}
         \centering
         \includegraphics[width=0.7\columnwidth]{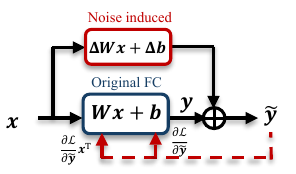}
         \caption{FC layer with noise injection.}
     \end{subfigure}     
     \begin{subfigure}{0.8\columnwidth}
         \centering
         \includegraphics[width=0.9\columnwidth]{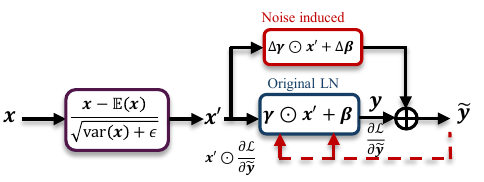}
         \caption{LN layer with noise injection.}
     \end{subfigure}

  \caption{The forward and backward processes for both FC and LN layers with noise injection.}
  \label{fig:fwd_backwd_nns}
\end{figure}

\subsection{Neural Network Weights Over-the-Air}\label{sec:OTA-NeRF}
In the previous subsection, we assume the weights, $\mathcal{W}$, of the OTA-NeRF model shown in Fig. \ref{fig:point_nerf_model} are perfectly available at the decoder side. 
This can be achieved by first applying lossless source coding to transform $\mathcal{W}$ into a bit sequence which is further channel coded and modulated to be transmitted over the channel. 
However, digital coded modulation suffers from cliff and leveling effects, which limit its performance when accurate channel estimation is no available. Therefore, we consider an alternative approach, namely, the OTA-NeRF framework, which is presented next.

As shown in Fig. \ref{fig:point_nerf_model}, the proposed OTA-NeRF model is comprised of two different types of layers, namely, the FC and the LN layers, whose input-output relationships are expressed as:
\begin{align}
    \bm{y} &= \bm{W}\bm{x} + \bm{b}
    \label{eq:input_output}
\end{align}
for the FC layer, and
\begin{align}
    \bm{y} &= \frac{\bm{x} - \mathbb{E}(\bm{x})}{\sqrt{\text{var}(\bm{x}) + \epsilon}} \odot \bm{\gamma} + \bm{\beta}
    \label{eq:input_output2}
\end{align}
for the LN layer. Here, $\text{var}(\bm{x})$ denotes the variance of the input vector $\bm{x}$, and $\epsilon$ is introduced for computational stability. We first assume the weights and the biases are transmitted over the AWGN channel (the algorithm is also applicable to fading channels). Take the FC layer as an example, the received signal at the receiver can be expressed as:
\begin{align}
    \widetilde{\bm{W}} &= \bm{W} + \Delta \bm{W}, \notag \\
    \tilde{\bm{b}} &= \bm{b} + \Delta \bm{b},
    \label{eq:noisy_fc}
\end{align}
where both $\Delta \bm{W}$ and $\Delta \bm{b}$ are the Guassian noise components added to the original weights and biases, $\bm{W}$ and $\bm{b}$, satisfying:
\begin{align}
    \|\bm{W}\|^2_F &=  \text{SNR} \; \mathbb{E}(\|\Delta \bm{W}\|^2_F), \notag \\
    \|\bm{b}\|^2_2 &=  \text{SNR} \; \mathbb{E}(\|\Delta \bm{b}\|^2_2).
    \label{eq:noise_power_weight}
\end{align}
The new input-output relationship of the FC layer with noise can be expressed as:
\begin{align}
    \tilde{\bm{y}} &= \underbrace{\bm{W}\bm{x} + \bm{b}}_{
    \text{original} \; \bm{y}} + \underbrace{(\Delta\bm{W}\bm{x} + \Delta\bm{b})}_{\text{noise induced}}.
    \label{eq:noisy_input_output}
\end{align}

The training objective of OTA-NeRF is to obtain $\bm{W}$ and $\bm{b}$ that are robust to different $\Delta \bm{W}$ and $\Delta \bm{b}$ realizations. The update rule of $\bm{W}$ and $\bm{b}$ can be expressed as:
\begin{align}
    \bm{W} &\leftarrow {\bm{W}} - \eta \frac{\partial \mathcal{L}}{\partial \tilde{\bm{y}}}\bm{x}^\top, \notag\\
    \bm{b} &\leftarrow {\bm{b}} - \eta \frac{\partial \mathcal{L}}{\partial \tilde{\bm{y}}},
    \label{eq:fc_noise_optim}
\end{align}
where $\eta$ is the learning rate. Note that compared with the original optimization procedure, where the gradient is $\frac{\partial \mathcal{L}}{\partial {\bm{y}}}$, we have $\frac{\partial \mathcal{L}}{\partial \tilde{\bm{y}}}$ due to the noise in the forward process.

The update rule for $\bm{\gamma}$ and $\bm{\beta}$ of the LN layer follows similarly to \eqref{eq:fc_noise_optim}:
\begin{align}
    \bm{\gamma} & \leftarrow \bm{\gamma} - \eta \bm{x}^\prime \odot \frac{\partial \mathcal{L}}{\partial \tilde{\bm{y}}}, \notag\\
    \bm{\beta} &\leftarrow \bm{\beta} - \eta \frac{\partial \mathcal{L}}{\partial \tilde{\bm{y}}},
    \label{eq:ln_noise_optim}
\end{align}
where $\bm{x}^\prime = \frac{\bm{x} - \mathbb{E}(\bm{x})}{\sqrt{\text{var}(\bm{x}) + \epsilon}}$.

The forward and backward processes of the noisy FC and LN layers during training are summarized in Fig. \ref{fig:fwd_backwd_nns}. 
After training, we obtain the neural network weights, $\mathcal{W}$, which will be transmitted to the receiver directly over the noisy channel without any further coding. The block occupancy information, $\mathcal{O}_B$, and the number of reconstructed points, $N_p$, on the other hand, are protected by ultra-reliable channel codes, and are assumed to be available at the decoder side. It is worth mentioning that, the $N_p$ value is determined via simulations at the transmitter analogous to that in \cite{deepwive}. In particular, for a specific $N_p$ value, the encoder calculates the average $D1$ performance for a large number of noise realizations, and selects the $N_p$ value that results in the best average $D1$ performance.
The training and testing processes of the proposed OTA-NeRF model are summarized in Algorithm \ref{algorithm:ota_point_nerf}.

  \begin{figure}[t]
\centering
\includegraphics[width=0.7\linewidth]{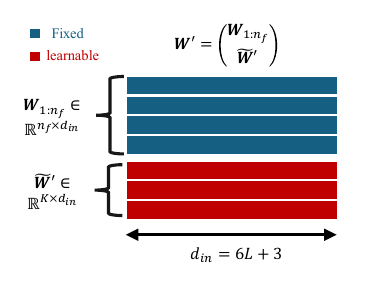}
\caption{We fix $n_f$ rows/elements of the weight ($\bm{W}$) / bias ($\bm{b}$) while update the remaining $K$ rows/elements to obtain the weight and bias, $\bm{W}^\prime$ and $\bm{b}^\prime$, for point cloud $\mathcal{P}^\prime$.}
\label{fig:finetune_weight}
\end{figure}

\subsection{Fine-tuning for Bandwidth Efficiency} \label{sec:finetune}
In the previous subsection, a neural network with parameters $\mathcal{W}$ is trained and transmitted over the noisy channel for a specific point cloud, denoted by $\mathcal{P}$. When a new point cloud, $\mathcal{P}^\prime$, arrives, a new neural network with parameters, $\mathcal{W}^\prime$, should be trained from scratch. Since the point clouds $\mathcal{P}$ and $\mathcal{P}^\prime$ might share some features, it is possible to transmit less number of parameters for $\mathcal{W}^\prime$  given that $\mathcal{W}$ is already available at the destination. In particular, we consider reducing the size and the number of training epochs to obtain $\mathcal{W}^\prime$ for $\mathcal{P}^\prime$ by fine-tuning only a small fraction of the weights of $\mathcal{W}$.

To start with, we emphasize that it is critical to determine which part of the parameters to freeze/fine-tune. For instance, one may update the parameters in the residual blocks while keeping the feature extraction and final classification modules fixed or one can fine-tune the parameters of the feature extraction module while leaving the remaining parameters frozen. We found through ablation experiments that it is more beneficial to update the parameters in the feature extraction and final classification modules, which consists of $d_{in}(d_1 + 1)$ and $(d_2+1)$ parameters, respectively. Since the number of parameters in the final classification module is relatively small, we will update all its $(d_2+1)$ elements. For the feature extraction module, on the other hand, we only fine-tune part of its FC layer. 

Let $\bm{W} \in \mathbb{R}^{d_{in} \times d_1}$ and $\bm{b} \in \mathbb{R}^{d_1}$ denote the weight and the bias of the FC layer,  respectively, and $K$ control the number of parameters for fine-tuning. For the new point cloud, $\mathcal{P}^\prime$, we fix the first $n_f \triangleq d_1 - K$ rows of $\bm{W}$ and the first $n_f$ elements of $\bm{b}$. The remaining $K$ rows and $K$ elements are replaced by the new values, $\widetilde{\bm{W}}^\prime \in \mathbb{R}^{K \times  d_{in}}$ and $\tilde{\bm{b}}^\prime \in \mathbb{R}^{K}$, respectively. The updated feature vector ${\bm{f}^\prime}$ for $\mathcal{P}^\prime$ can be expressed as:
\begin{align}
    \bm{f}^\prime = \begin{bmatrix}
    \bm{W}_{1:n_f}\tilde{\bm{x}} + \bm{b}_{1:n_f} \\
    \widetilde{\bm{W}}^\prime \tilde{\bm{x}} + \tilde{\bm{b}}^\prime \\
\end{bmatrix}.
    \label{eq:ft_feature}
\end{align}
The effective weight for the new point cloud is $\bm{W}^\prime = \begin{bmatrix} \bm{W}_{1:n_f} \\ \widetilde{\bm{W}}^\prime \end{bmatrix}$, while the effective bias is: $\bm{b}^\prime = \begin{bmatrix} \bm{b}_{1:n_f} \\ \widetilde{\bm{b}}^\prime \end{bmatrix}$. Both of them are shown in Fig. \ref{fig:finetune_weight} for a better illustration.

Since the parameters of the residual blocks are fixed, the number of real channel uses to transmit the neural network weights for $\mathcal{P}^\prime$ is simply $(d_{in} + 1)K + d_1 + 1$. By changing $K \in [0, d_1]$, different bandwidth and reconstruction performances can be obtained at the receiver. We will show through experiments that the proposed fine-tuning scheme is capable of achieving satisfactory reconstruction performance while using less channel bandwidth.

\begin{algorithm}
	\caption{Training and evaluation algorithm for the proposed OTA-NeRF framework.}
    \label{algorithm:ota_point_nerf}

	\SetKwInOut{Input}{Input}\SetKwInOut{Output}{Output}
	\SetKwFunction{Voxelization}{Voxelization}
	\SetKwFunction{BlockPartition}{BlockPartition}
 \SetKwFunction{OTANeRFForward}{OTANeRFForward}
  \SetKwFunction{IsFCLayer}{IsFCLayer}
  \SetKwFunction{GenerateVoxel}{GenerateVoxel}
\SetKwFunction{SimulateBestNp}{SimulateBestNp}
\SetKwFunction{PruneVoxel}{PruneVoxel}
 
	\Input{$\mathcal{P}, V,  B, L,  N_{epoch}, \eta, \mathrm{SNR}$}
	\Output{$\hat{\mathcal{P}}, \mathcal{W}$} 
	
	\BlankLine

        $\mathcal{G}^\prime \leftarrow \Voxelization(\mathcal{P}, V)$
        
        $\mathcal{G} \triangleq \{\mathcal{V}, \mathcal{O}, \mathcal{O}_B\} \leftarrow \BlockPartition(\mathcal{G}^\prime, B)$ \Comment{Point cloud preprocessing.}

       \%\% \textbf{Training Phase}.
       
	\For{$n=1$ \KwTo {$N_{epoch}$}}{
        \For{Each batch $\bm{x} \in \mathcal{V}$}{
            $\tilde{\bm{x}} = [\bm{x}, \sin(2^0 \pi \bm{x}), \cos(2^0 \pi \bm{x}), \ldots,  \sin(2^{L-1} \pi \bm{x}), $\\ $ \cos(2^{L-1} \pi \bm{x})],$  \Comment{Positional embedding.}
            
        $\widetilde{\mathcal{W}} \leftarrow
        \mathcal{W} + \Delta \mathcal{W}$ \Comment{Noisy channel.}
        
        $\bm{p} \leftarrow \OTANeRFForward(\{\mathcal{V}, \widetilde{\mathcal{W}}\})$.
        
        Calculate $
    \mathcal{L} = \frac{1}{N_{V}} \sum_{v = 1}^{N_{V}} - (o_v \log(p_v) + (1-o_v) \log(1-p_v))$.

    \For{weights  $ \in \mathcal{W}$}{
    \If{$\IsFCLayer(\text{weights})$}{
    Update $\{\bm{W}, \bm{b}\}$ via Equ. \eqref{eq:fc_noise_optim};
    }
    \Else{
        Update $\{\bm{\gamma}, \bm{\beta}\}$ via Equ. \eqref{eq:ln_noise_optim};
    }
        
		}}}
\%\% \textbf{Evaluation Phase}.

{$N_p \leftarrow \SimulateBestNp(\mathcal{W}, \mathrm{SNR})$.}

Transmit $\mathcal{W}$ with analog transmission and $\mathcal{O}_B, N_p$ with digital transmission.

$\mathcal{V} \leftarrow \GenerateVoxel(\mathcal{O}_B)$.

$\bm{p} \leftarrow \OTANeRFForward(\mathcal{V}, \hat{\mathcal{W}})$.

$\hat{\mathcal{P}} \leftarrow \PruneVoxel(\bm{p}, N_p)$.

	\BlankLine
\end{algorithm}

\section{{OTA-MetaNeRF framework}}\label{sec:VI}
\begin{figure}[t]
\centering
\includegraphics[width=\linewidth]{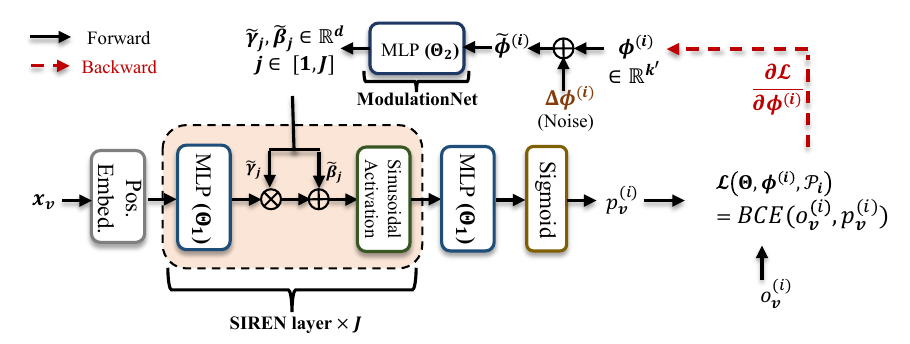}
\caption{{The illustration of the proposed OTA-MetaNeRF encoding process during the evaluation phase. The neural network weights, $\bm{\Theta} \triangleq \{\bm{\Theta}_1, \bm{\Theta}_2\}$, of the SIREN layers ($\bm{\Theta}_1$) and ModulationNet ($\bm{\Theta}_2$) are fixed, and only the latent vectors, $\bm{\phi}^{(i)}$, are obtained and transmitted for each input sample. The effect of the channel noise is taken into account when updating latent vectors, $\bm{\phi}^{(i)}$.}}
\label{fig:coinpp_illu}
\end{figure}  

{The OTA-NeRF framework presented in the previous section trains/fine-tunes neural network weights for each of the point cloud. 
In this section, we introduce a novel paradigm called OTA-MetaNeRF, which further extends the fine-tuning idea through meta-learning, and achieves more effective and efficient encoding.

\subsection{Representing Point Clouds via Latent Vectors}
Instead of representing each point cloud as neural network weights, we consider adopting multiple latent vectors to represent each point cloud. In particular, the same point cloud prepossessing procedure in Section \ref{sec:IV} is performed to obtain the discrete representation, $\mathcal{G} \triangleq \{\mathcal{V}, \mathcal{O}, \mathcal{O}_B\}$, as in \eqref{eq:discrete_pc}, consisting of $N_B$ occupied blocks. For each of the $b$-th, $b \in [1, N_B]$ block of the $i$-th point cloud, we assign a latent vector, $\bm{\phi}^{(i,b)} \in \mathbb{R}^{k'}$, to represent the block, which is detailed as follows.

Similarly to the OTA-NeRF framework, for each of the $b$-th block, we predict the occupancy of its $v$-th voxel, by feeding its coordinates, $\bm{x}_v$, into a neural network, termed as meta network, comprised of ModulationNet and SIREN layers parameterized by $\bm{\Theta}_1$ and $\bm{\Theta}_2$, respectively as shown in Fig. \ref{fig:coinpp_illu}. The overall parameters of the meta network is denoted by $\bm{\Theta} \triangleq \{\bm{\Theta}_1, \bm{\Theta}_2\}$.  The occupancy information for the $b$-th block corresponding to the $i$-th point cloud, is conveyed through latent vectors, $\bm{\phi}^{(i,b)}$. For simplicity, we first omit the block index $b$, and use $\bm{\phi}^{(i)}$ to represent an arbitrary block within the $i$-th point cloud.

The latent vector, $\bm{\phi}^{(i)}$ is fed into ModulationNet, comprised of a single MLP layer with dimensions $Jd \times k$, to obtain the modulations $\{\bm{\gamma}_j, \bm{\beta}_j\}, j\in [1, J]$:
\begin{equation}
    \{\bm{\gamma}_j, \bm{\beta}_j\}_{j\in [1, J]} \leftarrow \text{Modulate}(\bm{\phi}^{(i)}),
    \label{equ:modulation}
\end{equation}
where $\bm{\gamma}_j, \bm{\beta}_j \in \mathbb{R}^d$ are $d$-dimensional vectors, which will be used to modulate the hidden features in the SIREN layer, while $J$ denotes the number of SIREN layers \cite{coinpp}. Each SIREN layer is comprised of an MLP layer with sinusoidal activation function. The input coordinates, $\bm{x}_v$, is first positionally embedded according to \eqref{eq:pos_emb} to generate a higher dimensional vector $\tilde{\bm{x}}_v \in \mathbb{R}^{6L+3}$ for a richer representation, where $L$ denotes the number of frequencies. Then, the input and output relationship of the $j$-th SIREN layer with modulations can be expressed as:
\begin{equation}
    \bm{h}_{j+1} = \sin\left(\omega_0 (\bm{\gamma}_j \odot \text{MLP}(\bm{h}_{j}) +\bm{\beta}_j)\right),
    \label{equ:siren}
\end{equation}
where $\sin(\cdot)$ is the sinusoidal activation function, $\omega_0$ is the positive frequency value, $\bm{h}_{j}, \bm{h}_{j+1} \in \mathbb{R}^d$ denote the input and the output of the $j$-th modulated SIREN layer, respectively, and finally, $\text{MLP}(\cdot)$ denotes the MLP layer within the SIREN layer. Note that we initialize the input to the first SIREN layer using the positionally embedded vector, i.e., $\bm{h}_1 \leftarrow \tilde{\bm{x}}_v$.
After obtaining the modulated vector of the $J$-th SIREN layer, the final MLP layer with dimension $d$ is adopted followed by a sigmoid activation function to produce the predicted occupancy information as a probability value, $p_v^{(i)} \in (0, 1)$. 
It can be understood from \eqref{equ:siren} that the modulated vector $\bm{h}_{j+1}$ is a function of $\bm{\phi}^{(i)}$ through $\bm{\gamma}_j, \bm{\beta}_j$, which distinguishes its final occupancy prediction from other blocks with different latent vectors, $\bm{\phi}^{(k)}$.

\subsection{Training and Evaluation of OTA-MetaNeRF}
We then illustrate the training methodology of the proposed OTA-MetaNeRF. In particular, we are interested in optimizing the parameters ($\bm{\Theta}$) of the SIREN layers and the ModulationNet as well as the latent vector, $\bm{\phi}$.

\subsubsection{Training without noise}
For a better understanding, we start by assuming a perfect channel, i.e., the latent vector, $\bm{\phi}$, is perfectly available to the receiver.
We consider training with a batch size of $N_{bs}$. Note that different point clouds have different number of blocks, $N_B$, thus, to formulate the training batch, we first select $N_{bs}$ point clouds in the training dataset, and for each selected point cloud, we randomly sample a single block from its $N_B$ blocks.
Each data sample from the training batch contains $2^{3(V-B)}$ voxels and we denote the occupancy information of the $v$-th voxel belonging to the sampled block of the $i$-th point cloud, $\mathcal{P}_i$, as $o^{(i)}_v \in \{0, 1\}, v\in [1, 2^{3(V-B)}]$. After obtaining the final probability output as explained in the previous subsection, the corresponding loss function can be calculated as:
\begin{align}
    \mathcal{L}(\bm{\Theta, \bm{\phi}}^{(i)}, \mathcal{P}_i) &=    \sum_{v = 1}^{N_{V}} - o^{(i)}_v \log(p^{(i)}_v) \notag \\ 
    &- (1-o^{(i)}_v) \log(1-p^{(i)}_v).
    \label{eq:coin_bce_loss}
\end{align}

We then illustrate the optimization process of the neural network parameters, $\bm{\Theta}$, and latent vectors, $\bm{\phi}^{(i)}$. In particular, for each of the $i$-th point cloud, we first perform $T$-step ($T\ge 1$) gradient descent w.r.t. $\bm{\phi}^{(i)}$. For the $t$-th step ($t\in [1, T]$), the latent vector, $\bm{\phi}^{(i)}_t$, is updated as:
\begin{equation}
    \bm{\phi}^{(i)}_t \leftarrow \bm{\phi}^{(i)}_{t-1} - \alpha \nabla_{\bm{\phi}} \mathcal{L}(\bm{\Theta}, \bm{\phi}^{(i)}_{t-1}, \mathcal{P}_i),
    \label{eq:gd_phi}
\end{equation}
where $\alpha$ denotes the learning rate to optimize the latent vector and the loss function is calculated for each step using the updated latent vector from the previous step. Furthermore, we note that it suffices to initialize $\bm{\phi}^{(i)}_0$ using an all-zero vector, i.e., $\bm{\phi}^{(i)}_0 \leftarrow \bm{0}_{k'}$.

Different with the optimization of the latent vector, $\bm{\phi}^{(i)}, i \in [1, N_{bs}]$, which is specific for the $i$-th point cloud,  $\bm{\Theta}$ parameters of the meta network are obtained through meta-learning for all point cloud inputs, through the following gradient steps:
\begin{equation}
    \bm{\Theta} \leftarrow \bm{\Theta} - \beta \sum_{i=1}^{N_{bs}} \nabla_{\bm{\Theta}} \mathcal{L}(\bm{\Theta}, \bm{\phi}^{(i)}_{T}, \mathcal{P}_i),
    \label{eq:gd_theta}
\end{equation}
where $\bm{\phi}^{(i)}_{T}$ denotes the optimized latent vector after $T$ steps as in 
\eqref{eq:gd_phi} corresponding to the $i$-th point cloud, and $\beta$ is the meta-learning rate for the meta network.
Finally, the model is trained by alternatively optimizing the latent vectors and the neural network weights of the meta network parameters over the training dataset.

\begin{algorithm}
	\caption{Training and evaluation algorithm for the proposed OTA-MetaNeRF framework.}
    \label{algorithm:ota_metanerf}

	\SetKwInOut{Input}{Input}\SetKwInOut{Output}{Output}
	\SetKwFunction{Voxelization}{Voxelization}
	\SetKwFunction{BlockPartition}{BlockPartition}
    \SetKwFunction{Modulate}{Modulate}
    \SetKwFunction{SIRENForward}{SIRENForward}
    \SetKwFunction{GetOccupancyLoss}{GetOccupancyLoss}
    \SetKwFunction{UpdatePhi}{UpdatePhi}
    \SetKwFunction{UpdateTheta}{UpdateTheta}
    \SetKwFunction{SimulateBestNp}{SimulateBestNp}
    \SetKwFunction{TransmitPhi}{TransmitPhi}
    \SetKwFunction{PredictOccupancy}{PredictOccupancy}
    \SetKwFunction{ConstructPointCloud}{ConstructPointCloud}
 
	\Input{$\{\mathcal{P}_{train}\}, \{\mathcal{P}_{test}\}, V, B, L, k', J, N_{epoch}, T, \alpha, \beta, \mathrm{SNR}$}
	\Output{$\{\hat{\mathcal{P}}_{test}\}$} 
	
	\BlankLine
        
    

    Preprocess all point clouds according to line 1-2 in Algorithm \ref{algorithm:ota_point_nerf}.

    \%\% \textbf{Training Phase}.
       
	\For{$n=1$ \KwTo {$N_{epoch}$}}{
            \For{$i$-th point cloud $\mathcal{P}_i \in \{\mathcal{P}_{train}\}$}{
            Randomly select one block with occupancy $\bm{o}^{(i)}$.
            Initialize latent vector: $\bm{\phi}^{(i)}_0 \leftarrow \bm{0}_{k'}$.
	    
	    \For{$t = 1$ \KwTo $T$}{
	        
	        $\bm{\tilde{\phi}}^{(i)}_{t-1} \leftarrow \bm{\phi}^{(i)}_{t-1} + \Delta \bm{\phi}^{(i)}_{t-1}$. \Comment{Add noise \eqref{eq:noise_power_phi}.}
	        
	        $\{\bm{\gamma}_j, \bm{\beta}_j\}_{j=1}^J \leftarrow \Modulate(\bm{\tilde{\phi}}^{(i)}_{t-1})$.
	        
	        $\bm{p}^{(i)} \leftarrow \SIRENForward(\bm{x}_v, \{\bm{\gamma}_j, \bm{\beta}_j\})$.
	        
	        $\mathcal{L}_i \leftarrow  \sum_{v = 1}^{N_{V}} - o^{(i)}_v \log(p^{(i)}_v) - (1-o^{(i)}_v) \log(1-p^{(i)}_v)$.
	        
	        $\bm{\phi}^{(i)}_t \leftarrow \UpdatePhi(\bm{\phi}^{(i)}_{t-1}, \mathcal{L}, \alpha)$. \Comment{Eqn. \eqref{eq:gd_noisy_phi}}
	    }
        $\bm{\Theta} \leftarrow \UpdateTheta(\bm{\Theta}, \mathcal{L}_i, \beta)$. \Comment{Eqn. \eqref{eq:gd_noisy_theta}}
            }

	}

	\%\% \textbf{Evaluation Phase}.
	
	\For{each point cloud $\mathcal{P} \in \{\mathcal{P}_{test}\}$}{
	
	    Transmit $\mathcal{O}_B$ digitally.
        
	    \For{$b \in [1, N_B]$}{
	    Initialize latent vector: $\bm{\phi}^{(b)}_0 \leftarrow \bm{0}_{k'}$.
	    
	    \For{$t = 1$ \KwTo $T$}{
	        $\bm{\tilde{\phi}}^{(b)}_{t-1} \leftarrow \bm{\phi}^{(b)}_{t-1} + \Delta \bm{\phi}^{(b)}_{t-1}$.
	        
	        Update latent vector: $\bm{\phi}^{(b)}_t \leftarrow \UpdatePhi(\bm{\phi}^{(b)}_{t-1}, \mathcal{L}, \alpha)$
	    }
	    
	    Transmit $\bm{\phi}^{(b)}_T$ using analog transmission.
	    
	    $\bm{p}^{(b)} \leftarrow \text{Repeat line 8 \& 9 using} \, \bm{\tilde{\phi}}^{(b)}_T$ and $\bm{\Theta}$}

	$\hat{\mathcal{P}} \leftarrow \ConstructPointCloud(\{\bm{p}^{(b)}\}_{b=1}^{N_B}, \mathcal{O}_B)$}

	\BlankLine
\end{algorithm}

\subsubsection{Training with noise}
Next, we introduce the training process which takes the channel noise into account. For the $i$-th point cloud, the received signal is the noisy version of the optimized latent vector\footnote{{We omit the subscript, $T$, of the optimized latent vector for simplicity.}}, $\bm{\phi}^{(i)}$:
\begin{equation}
    \bm{\tilde{\phi}}^{(i)} = \bm{{\phi}}^{(i)} + \Delta\bm{{\phi}}^{(i)},
    \label{eq:noisy_phi}
\end{equation}
where $\Delta\bm{{\phi}}^{(i)}$, as in \eqref{eq:noise_power_weight}, denotes the noise term satisfying:
\begin{align}
    \|\bm{{\phi}}^{(i)}\|^2_2 &=  \text{SNR} \; \mathbb{E}(\|\Delta \bm{{\phi}}^{(i)}\|^2_2).
    \label{eq:noise_power_phi}
\end{align}

The noise introduced by the wireless channel degrades the modulations $\{\bm{\gamma}_j, \bm{\beta}_j\}$ in \eqref{equ:modulation} into a noisy version, $\{\tilde{\bm{\gamma}}_j, \tilde{\bm{\beta}}_j\}$, leading to less satisfactory occupancy prediction performance. To mitigate this, we train the meta network and the latent vectors taking the noise term into account.  The optimization methods follow similarly to the noiseless scenario, yet the update rules of the parameters change as follows:
\begin{subequations} 
\begin{align}
    \bm{\phi}^{(i)}_t &\leftarrow \bm{\phi}^{(i)}_{t-1} - \alpha \nabla_{\bm{\phi}} \mathcal{L}(\bm{\Theta, \tilde{\bm{\phi}}}^{(i)}_{t-1}, \mathcal{P}_i), \quad t \in [1, T] \label{eq:gd_noisy_phi} \\
    \bm{\Theta} &\leftarrow \bm{\Theta} - \beta \sum_{i=1}^{N_{bs}} \nabla_{\bm{\Theta}} \mathcal{L}(\bm{\Theta, \tilde{\bm{\phi}}}^{(i)}_{T}, \mathcal{P}_i).
    \label{eq:gd_noisy_theta}
\end{align}
\end{subequations} 
It is worth emphasizing that we calculate the gradient of the loss function in \eqref{eq:gd_noisy_phi} w.r.t. the original latent vector, $\bm{\phi}^{(i)}$, instead of its noisy version, $\tilde{\bm{\phi}}^{(i)}$. Furthermore, we change the noise power according to the norm of the latent vector, $\|\bm{\phi}^{(i)}_t\|_2$,  which is updated during the optimization process.

\subsubsection{Evaluation phase}
Finally, with the trained OTA-MetaNeRF model, we reconstruct the new point cloud in the test dataset as follows. As in Section \ref{sec:IV}, the block occupancy information, $\mathcal{O}_B$, is assumed to be available at the receiver using digital transmission, and thus, only $N_B$ occupied blocks of the point cloud need to be reconstructed, reducing the reconstruction complexity for real-time point cloud transmission. The neural network parameters, $\bm{\Theta}$, of the meta network is assumed to be available at the transceiver; hence, the transmitter only needs to generate and transmit the latent vectors for the occupied blocks by performing \eqref{eq:gd_noisy_phi} $T$ times.
This leads to an average number of complex channel uses as:
\begin{equation}
    k \triangleq \mathbb{E}_{\mathcal{P}}\left(N_Bk^\prime\right)/2,
    \label{eq:bw_coinpp}
\end{equation}
where the expectation is taken over all point clouds and division by 2 is due to the compelx channel symbols.
Similarly to the reconstruction process for OTA-NeRF in Section \ref{sec:IV}, we determine the number of reconstructed points for each point cloud via simulations, which is transmitted as side information.

The optimized latent vector for each block is then transmitted to the receiver over the wireless channel and the receiver utilizes the noisy received signal to produce the prediction of the occupancy information of each block according to Section \ref{sec:IV}. The prediction output of the $N_B$ occupied blocks are organized to generate the final reconstruction of the point cloud. We summarize the proposed OTA-MetaNeRF framework in Algorithm \ref{algorithm:ota_metanerf} for a better understanding.
}

\section{Numerical Experiments}\label{sec:V}
This section presents the results of our numerical experiments to evaluate the reconstruction performance of the proposed SEPT, the OTA-NeRF and {OTA-MetaNeRF} frameworks. {We also provide the computational and run-time complexities of the proposed schemes to confirm their capabilities for real-time point cloud communications\footnote{{Source code is available at \url{https://github.com/aprilbian/OTA-PCT}.}}.}

\begin{figure*}
     \centering
     \begin{subfigure}{{0.66\columnwidth}}
         \centering
         \includegraphics[width=\columnwidth]{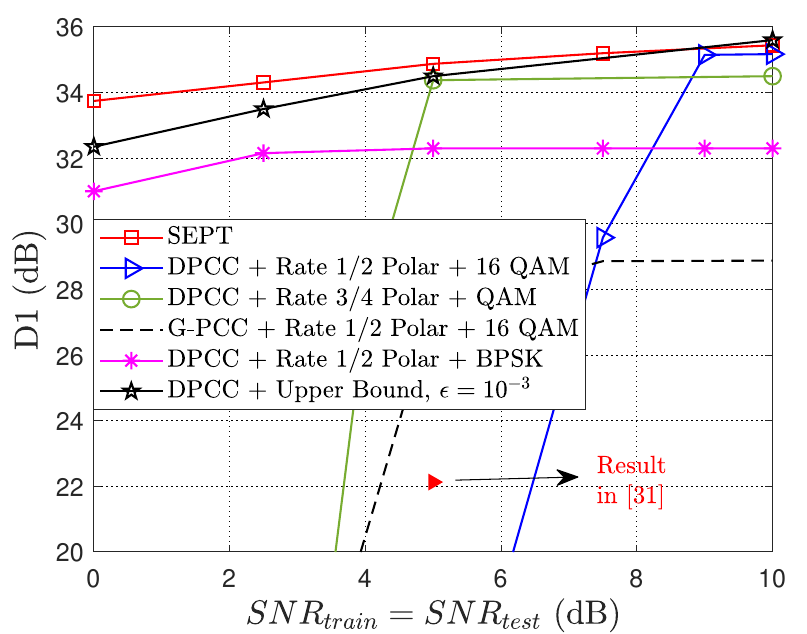}
         \caption{}
     \end{subfigure}
     \begin{subfigure}{{0.66\columnwidth}}
         \centering
         \includegraphics[width=0.96\columnwidth]{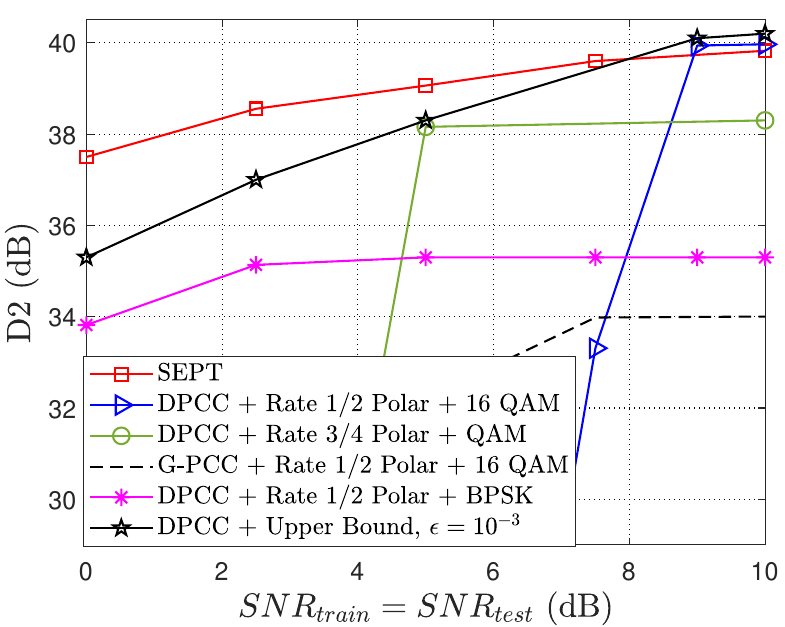}
         \caption{}
     \end{subfigure}
     \begin{subfigure}{0.66\columnwidth}
         \centering
         \includegraphics[width=\columnwidth]{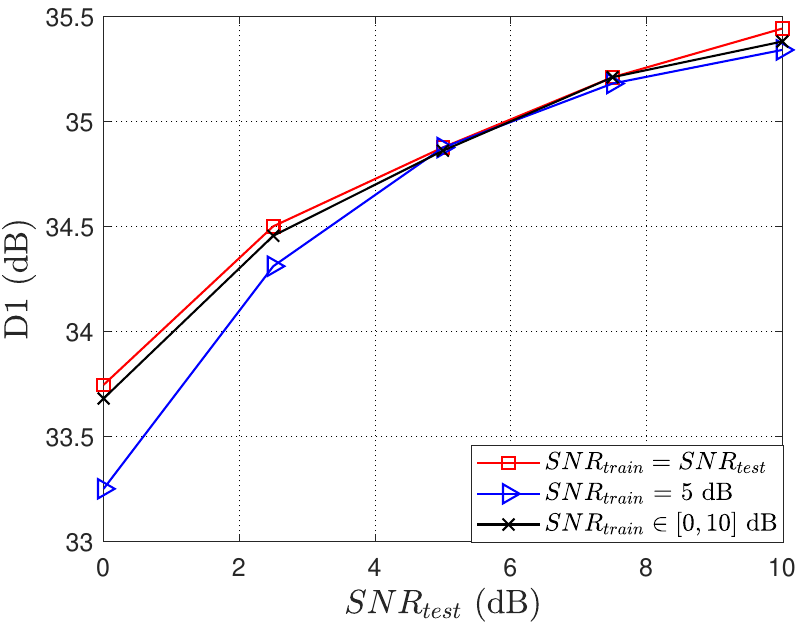}
         \caption{}
     \end{subfigure}
  \caption{Performance comparison over an AWGN channel: (a) $\&$ (b) the $\text{D1}$ and $\text{D2}$ performances, where $n = 100$; (c) {$\text{D1}$ versus $\mathrm{SNR}_{test}$ obtained by the SNR-adaptive SEPT model trained with $\mathrm{SNR}_{\text{train}} \in [0, 10]$ dB and $n = 100$.}}
\label{fig:sept_final_simu}
\end{figure*}

\subsection{Evaluation of SEPT}
SEPT is proposed mainly for small point clouds and we consider transmitting the point cloud data from the downsampled ShapeNet dataset \cite{chang2015shapenet}, which contains about $51000$ different shapes, and each point cloud is sampled to $N = 2048$ points using the FPS algorithm.
In both the SEPT encoder and decoder, the dimension of the intermediate attributes is set to $d_f = 256$ and the number of neurons in the MLPs of the coordinate reconstruction layer is set to $128$.
During training, we adopt the Adam optimizer with a varying learning rate, which is initialized to 0.001 and reduced by a factor of $0.5$ every $20$ epochs. 
We set the number of epochs to $200$ and the batch size to $32$.

\subsubsection{The reconstruction performance}
We first evaluate the reconstruction performance of SEPT with various channel SNR values over the AWGN channel. 
Two separate source-channel coding schemes are considered as benchmarks.
For source coding, the first benchmark uses the standard octree-based point cloud compression scheme, i.e., G-PCC.
The second benchmark, named DPCC, uses the state-of-the-art deep learning-based point cloud compression scheme with a focus on ultra low rate point cloud compression,  \cite{pct_pcc}.
Both schemes are protected by Polar codes with rate $\{1/2, 3/4\}$ and modulated by BPSK, QPSK, or 16QAM for transmission. {}We also provide the results for the DPCC delivered at the finite block length converse bound \cite{finite_length_capacity} for a block error rate of $\epsilon = 10^{-3}$.

\begin{figure}[t]
\centering
\includegraphics[width=0.8\linewidth]{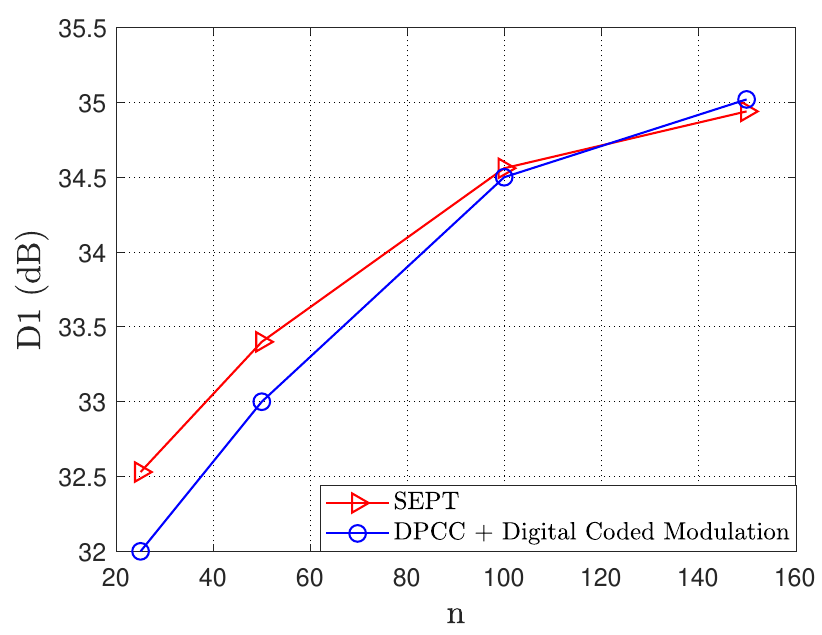}
\caption{The $\text{D1}$ performance of the proposed SEPT w.r.t. different number of channel uses $n$ over a Rayleigh fading channel with an average SNR of $10$ dB.}
\label{fig:varying_n}
\end{figure}

The simulation results are presented in Fig. \ref{fig:sept_final_simu}, where we fix the channel bandwidth to $n = 100$.
In Figs. \ref{fig:sept_final_simu}(a) and (b), a specific SEPT model is trained for each channel SNR value.
As shown, SEPT is significantly better than the separation based scheme with G-PCC. {In particular, to obtain the results for G-PCC, we use an average of $n = 430$ in the simulations. Despite the much larger channel bandwidth compared with that used by SEPT, G-PCC is still much worse in terms of $D1$ performance. This observation is also reported in \cite{pct_pcc}.}
SEPT also outperforms the separation-based scheme with DPCC \cite{pct_pcc}, especially in the low-SNR regime.
Note that DPCC can approach the performance of SEPT at high SNRs, if proper code rate and modulation schemes are selected. We observe that the performance of the separation-based scheme with DPCC falls short of the performance of SEPT at low SNR values even if we consider the finite-rate upper bound from \cite{finite_length_capacity}. This can be considered as the best performance separation-based schemes can achieve in combination with state-of-the-art point cloud compression scheme DPCC. {Another deep learning based point cloud transmission scheme is proposed in \cite{gnn_pct}, where a graph neural network (GNN) is adopted as the backbone to extract the features from the original point cloud at the transmitter, while simple FC layers are employed for point cloud reconstruction at the receiver. {We reproduce the results} in \cite{gnn_pct} and for the scenario with $n = 100$ and $\mathrm{SNR} = 5$ dB, it reports an average $D1$ value of 22.13 dB, whereas SEPT achieves 34.88 dB.}

Next, we illustrate the benefits of the proposed SA blocks for SNR-adaptive transmission.  In this simulation, the SNR-adaptive SEPT model is trained with $\mathrm{SNR}_\text{train} \in [0, 10]$ dB and tested under $\mathrm{SNR}_\text{test} = \{0,2.5,5,7.5,10\}$ dB. As shown in Fig. \ref{fig:sept_final_simu}(c), the SNR-adaptive SEPT model can achieve nearly the same performance with models trained and tested at the same SNR, showing the effectiveness of the proposed SA blocks. We also observe that, even though when tested at $\mathrm{SNR}_{test} = 0$ dB, the model trained at $\mathrm{SNR}_\text{train} = 5$ dB achieves an average $D1$ value approximately $0.5$ dB below that of the model trained under $\mathrm{SNR}_\text{train} = \mathrm{SNR}_\text{test}$, unlike digital coding approaches, the former model avoids the cliff and leveling effects when evaluated at different SNR values.

We then show the effectiveness of the SEPT over the Rayleigh fading channel as shown in Fig. \ref{fig:varying_n}. In this simulation, CSI $h$ is sampled from a complex Gaussian distribution, $\mathcal{CN}(0, 1)$, and the average channel SNR is fixed at $10$ dB. We evaluate the performance of the proposed SEPT w.r.t. different number of channel uses, $n$.  The digital baseline in this case uses DPCC \cite{pct_pcc} for compression and a combination of coded modulation schemes for transmission. To be specific, we search over different combinations of 64QAM + 1/2 Polar, 16QAM + 1/2 Polar, 4QAM + 1/2 Polar and BPSK + 1/2 Polar, and use the option with the best performance at each channel realization. We can observe that the average ${D1}$ performance of SEPT outperforms the digital baseline by 0.5 dB in the short block length regime of $n = 25$. The two schemes perform similarly, and they both start to saturate beyond $n \ge 100$ due to the fact that the max pooling operation at the transmitter focuses more on the global features while the fine details may be lost.

\begin{figure}
     \centering
     \begin{subfigure}{0.3\columnwidth}
         \centering
         \includegraphics[width=0.8\columnwidth]{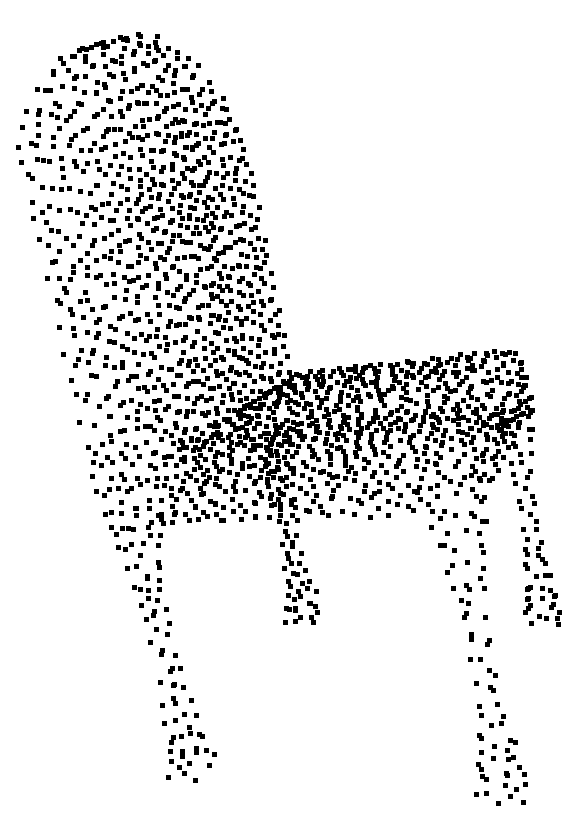}
     \end{subfigure}
     \begin{subfigure}{0.3\columnwidth}
         \centering
         \includegraphics[width=0.8\columnwidth]{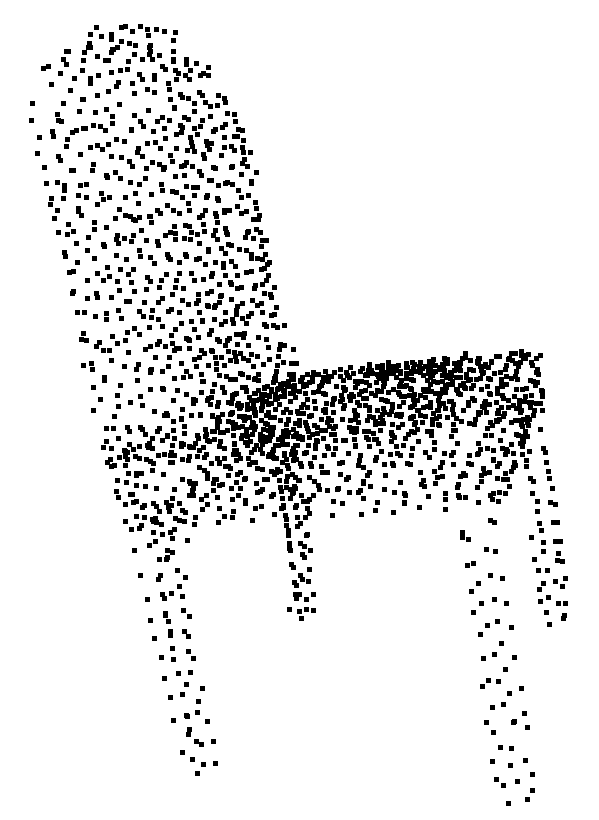}
     \end{subfigure}
     \begin{subfigure}{0.3\columnwidth}
         \centering
         \includegraphics[width=0.8\columnwidth]{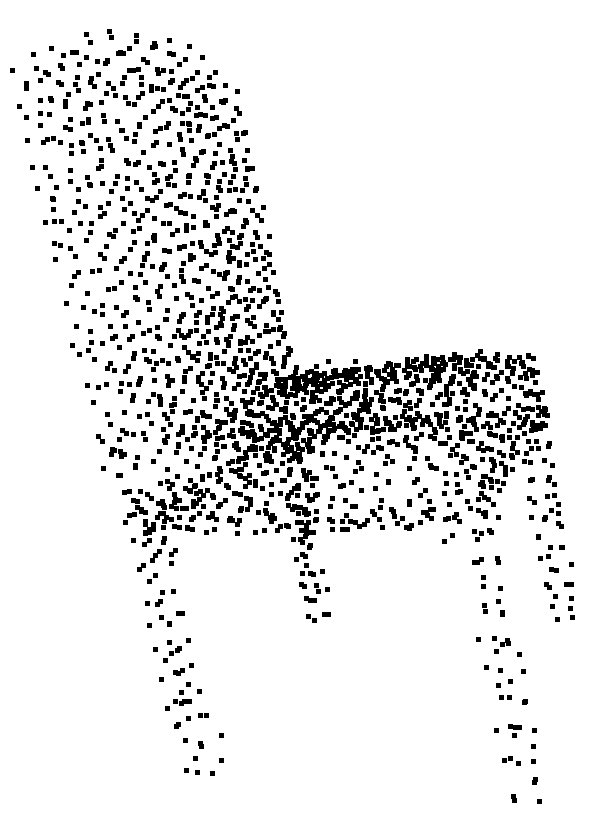}
     \end{subfigure}
  \caption{Visualization of reconstructed point clouds. Left: original $\mathcal{P}$; middle: $\hat{\mathcal{P}}$ with $n = 150, \mathrm{SNR} = 5$ dB; right: $\hat{\mathcal{P}}$ with $n = 150, \mathrm{SNR} = 0$ dB. }
\label{fig:visualize}
\end{figure}

We also provide a visualization of the reconstructed point clouds with $n = 150$ and $\mathrm{SNR} \in \{0, 5\}$ dB. As shown in Fig. \ref{fig:visualize}, SEPT yields visually pleasing results even when the $\mathrm{SNR}$ is as low as 0 dB.

{
\subsubsection{Complexity analysis of SEPT} Here, we provide the computational (big-$\mathcal{O}$) complexity as well as the run-time complexity of the SEPT scheme.
For a better understanding of the SEPT's scalability w.r.t. the number of points within the point clouds, we first provide the big-$\mathcal{O}$ complexity of the downsampling, self-attention and upsampling modules in Table \ref{tab:big_O}, where $N, k, d_f$ denote the number of input points, the number of neighbors in the kNN algorithm and the feature dimension, respectively. Moreover, the downsampling module employs $I$ 2D Convolutional layers whose filter size and stride are one while the input and output dimensions of each convolutional layer equal to $d_{f}$. It can be seen that the big-$\mathcal{O}$ complexities of different modules are linear w.r.t. the number of points, $N$, showing the proposed SEPT model is applicable for real-time applications.

Then, the running time for each SEPT module, i.e., downsampling, self-attention and upsampling modules are provided and the overall time cost is compare with the baseline \cite{pct_pcc} in Table \ref{tab:combined_complexity}. In this simulation, we set $n = 100, \mathrm{SNR} = 10$ dB.  The run-time complexities of the two schemes are evaluated on the same RTX 3080 GPU and is averaged over all point clouds in the test dataset. It can be seen that the SEPT scheme achieves much lower latency compared with the baseline. This is due to the fact that the baseline involves probability modeling and entropy coding of the latent vectors. Our SEPT, on the other hand, directly maps the point clouds to the channel codewords with less delay. 
}

\begin{table}[htbp]
\centering
\caption{{Computational and run-time complexities of the downsampling (Down.), self-attention (Attn.) and upsampling (Up.) modules for the SEPT model}.}
{ 
\begin{subtable}[t]{\linewidth}
    \centering
    
    \begin{tabular}{c|c|c|c}
        \hline
        Modules & Down. & Attn. & Up. \\
        \hline
        big-$\mathcal{O}$ complexity & $\mathcal{O}( k N I d^2_{f})$ & $\mathcal{O}( k N d^2_{f})$ & $\mathcal{O}(Nd^2_{f})$ \\
        \hline
    \end{tabular}
    \vspace{0.2cm}
    \caption{{Computational complexities of the modules.}}
    \label{tab:big_O}
\end{subtable}

\vspace{0.5em}

\begin{subtable}[t]{\linewidth}
    \centering
    
    \begin{tabular}{|c|cc|cc|c|c|}
        \hline
        \multirow{2}{*}{} & \multicolumn{2}{c|}{SEPT Encode} & \multicolumn{2}{c|}{SEPT Decode} & \multirow{2}{*}{Total} & \multirow{2}{*}{Baseline\cite{pct_pcc}} \\
        \cline{2-5}
        & Down. & \multicolumn{1}{|c|}{Attn.} & Refn. & \multicolumn{1}{|c|}{Up.} & & \\
        \hline
        Time (s) & 0.13 & \multicolumn{1}{|c|}{0.004} & 0.003 & \multicolumn{1}{|c|}{0.001} & \textbf{0.14} & 0.56 \\
        \hline
    \end{tabular}
    \vspace{0.2cm}
    \caption{{Run-time complexities of the SEPT model and the baseline scheme.}}
    \label{tab:sept_complex}
\end{subtable}
} 

\label{tab:combined_complexity}
\end{table}

\subsection{{Ablation Studies of the OTA-NeRF}}
We then evaluate the reconstruction performance of the proposed OTA-NeRF scheme. 
Here, we consider the Semantic KITTI dataset \cite{chang2015shapenet}, which is a large-scale outdoor-scene dataset for point cloud semantic segmentation. 

Different combinations of the number of frequencies, $L$, hidden neurons, $d_1, d_2$, and the residual blocks, $N_r$, are adopted to achieve different number of channel uses and reconstruction performances.
The Adam optimizer with a varying learning rate is adopted where the learning rate is initialized to $10^{-3}$ and decreases by $0.1$ every $30$ epochs.  The OTA-NeRF model is trained for $100$ epochs with a batch size of $4096$.

\begin{figure}[t]
\centering
\includegraphics[width=0.8\linewidth]{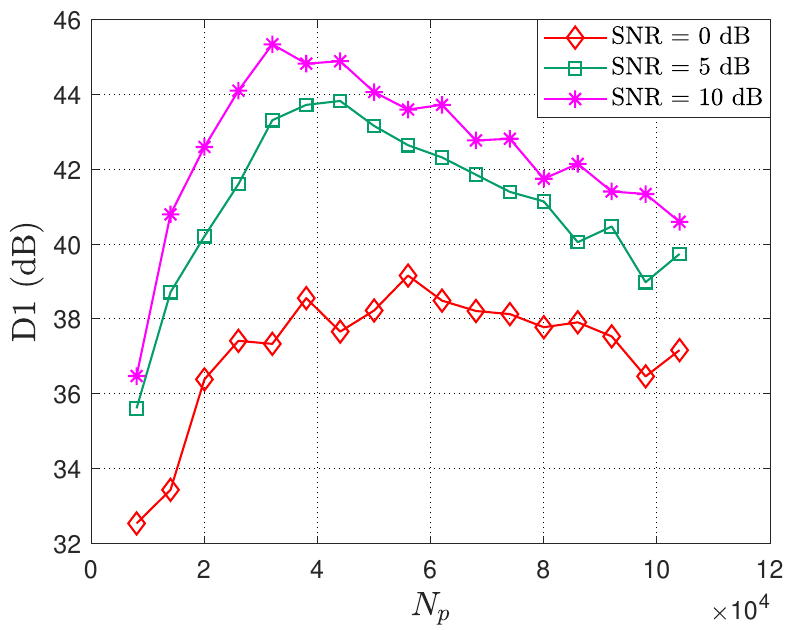}
\caption{The average $D1$ performance of OTA-NeRF w.r.t. different number of reconstructed points, $N_p$.}
\label{fig:optimal_Np}
\end{figure}

\begin{figure}[t]
\centering
\includegraphics[width=0.8\linewidth]{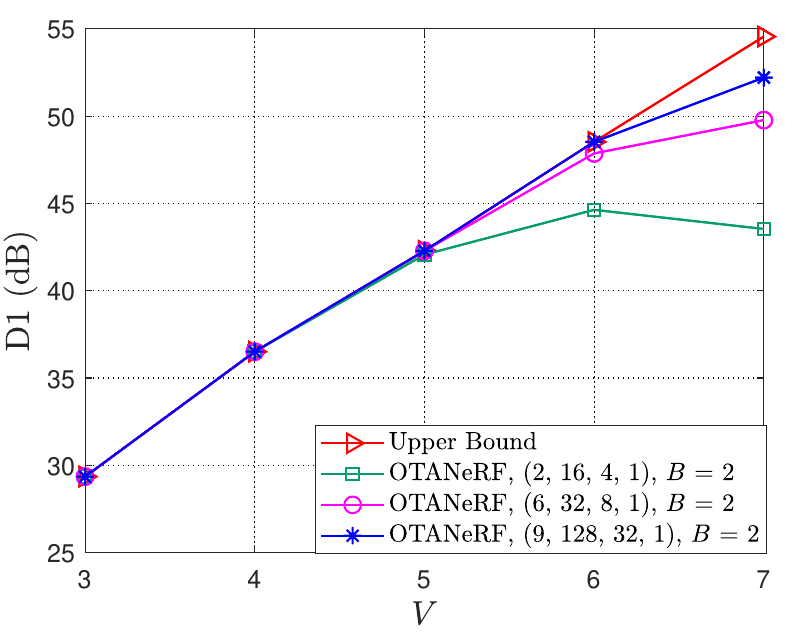}
\caption{The $D1$ performance w.r.t different voxel resolutions, $V$ with fixed block resolution, $B = 2$.}
\label{fig:num_voxels}
\end{figure}

\subsubsection{Effects of the number of reconstructed points}
We first illustrate that the number of the reconstructed points, $N_p$, plays an important role in achieving satisfactory reconstruction performance. In this simulation, we set the neural network parameters, $(L, d_1, d_2, N_r) = (6, 32, 8, 1)$. The numbers of voxels and blocks per dimension are set to $(2^V, 2^B) = 128, 4$, respectively, which leads to a discrete representation, $\mathcal{G}$ with the numbers of occupied voxels and blocks equal to $(N_o, N_B) = (9363, 23)$, respectively.  We consider an AWGN channel with $\mathrm{SNR} = \{0, 5, 10\}$ dB and the OTA-NeRF model is trained and evaluated on the first point cloud from the dataset as an example.
The average $D1$ performance w.r.t. $N_p$ is shown in Fig. \ref{fig:optimal_Np}. 

As can be seen, in the considered scenario, having $N_p > N_o$ yields better performance. It is also shown that when $\mathrm{SNR}$ is higher, e.g., $\mathrm{SNR} = 5, 10$ dB, a smaller $N_p$ is preferable. When $\mathrm{SNR} = 0$ dB, on the other hand, a larger $N_p$ leads to a better reconstruction performance. This aligns with the intuition that, when $\mathrm{SNR}$ is high, the occupancy of each voxel can be predicted more accurately and there is no need to reconstruct the points with relatively low probability values. As mentioned in the previous section, the optimal $N_p$ value is figured out at the transmitter and is transmitted digitally along with the block occupancy information, $\mathcal{O}_B$, as metadata to the destination for reconstruction.

\begin{figure*}
     \centering
     \begin{subfigure}{{0.64\columnwidth}}
         \centering
         \includegraphics[width=\columnwidth]{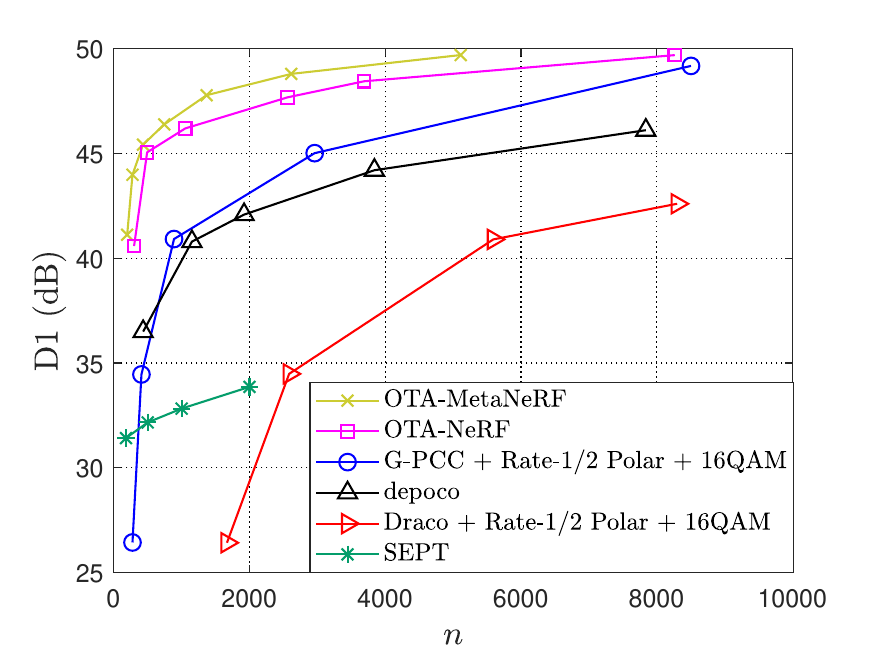}
         \caption{}
     \end{subfigure}
     \begin{subfigure}{{0.64\columnwidth}}
         \centering
         \includegraphics[width=\columnwidth]{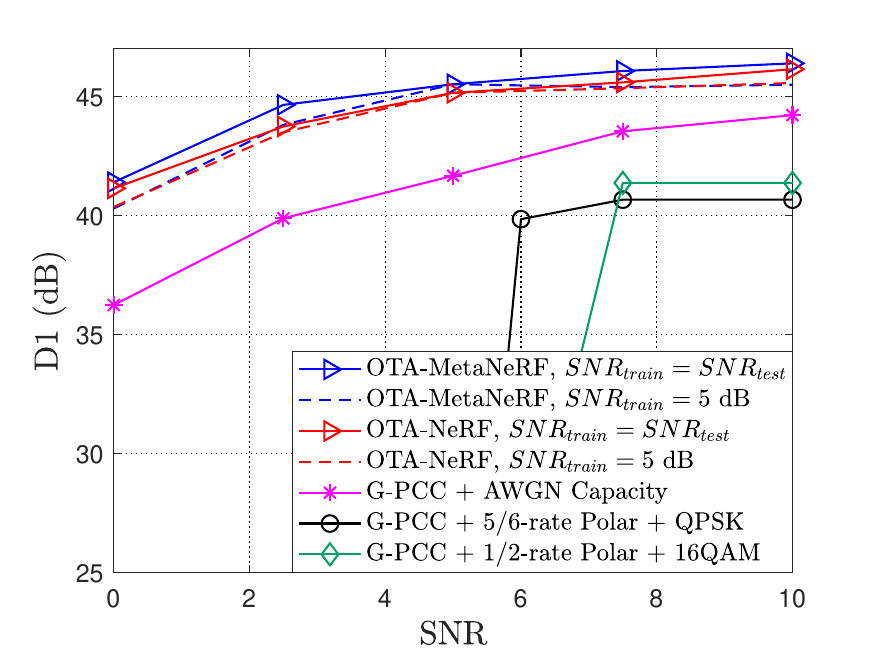}
         \caption{}
     \end{subfigure}
     \begin{subfigure}{0.64\columnwidth}
         \centering
         \includegraphics[width=\columnwidth]{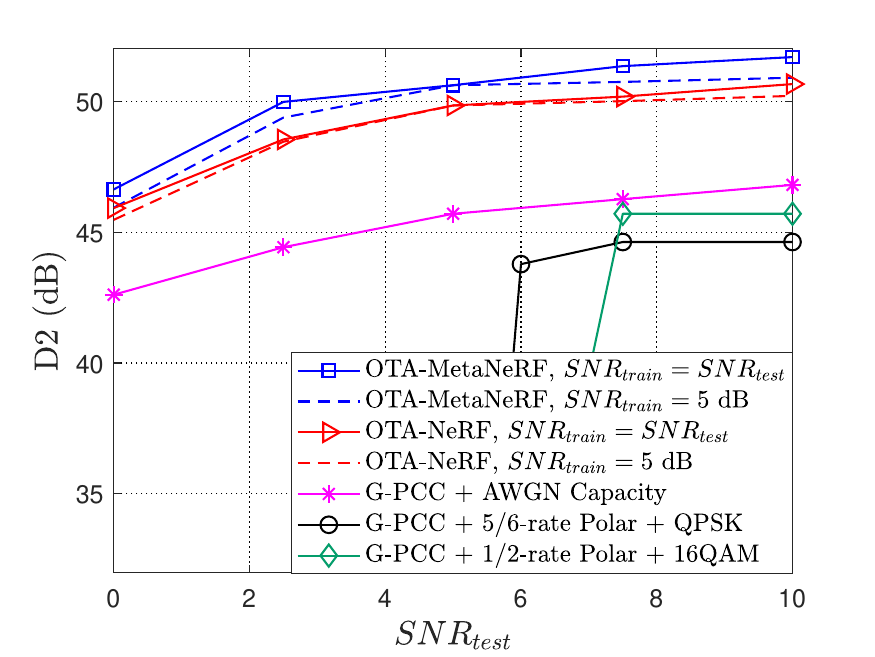}
         \caption{}
     \end{subfigure}
  \caption{{Reconstruction performances of the proposed OTA-NeRF and OTA-MetaNeRF schemes over the AWGN channel:  (a) $\text{D1}$ obtained by the OTA schemes and various baseline schemes with different numbers of complex channel uses, $n$, under $\mathrm{SNR} = 10$ dB; (b) $\&$ (c) $\text{D1}$ and $\text{D2}$ performances for different channel SNRs with a fixed number of complex channel uses.}}
\label{fig:final_simu}
\end{figure*}

\subsubsection{Voxel and block resolution}
We then explore the effects on the voxel and block resolutions, i.e., $V$ and $B$, respectively. In this simulation, we fix the block resolution $B = 2$ and vary $V \in [3, 7]$. Different OTA-NeRF models are trained and tested assuming noiseless channel with parameters $(L, d_1, d_2, N_r) = (2, 16, 4, 1)$, $(6, 32, 8, 1)$ and $(9, 128, 32, 1)$. Their $D1$ performances w.r.t different $V$ are shown in Fig. \ref{fig:num_voxels}.

As can be seen, the `Upper Bound' curve calculates the $D1$ performance between different discrete representation, $\mathcal{G}$, (with different $V$) and the original point cloud $\mathcal{P}$. It is shown that the performance improves with $V$, which aligns with the intuition that when $V$ is small, the total number of voxels decreases leading to substantial loss of details in the original point cloud. When $V$ becomes larger, the original point cloud information is well preserved leading to a higher $D1$ value. 
We can also observe from the figure that for the OTA-NeRF model with parameters $(L, d_1, d_2, N_r) = (2, 16, 4, 1)$, its reconstruction performance first improves when $V \in \{3, 4, 5, 6\}$ and then degrades when $V = 7$. This is due to the fact that the number of voxels, $|\mathcal{V}|$, of the occupied blocks grows exponentially w.r.t $V$, and it is impossible for a neural network with a limited number of parameters to predict the occupancy of all the voxels correctly. This problem can be resolved by increasing the number of neural network parameters as the $D1$ performances of the OTA-NeRF models with $(L, d_1, d_2, N_r) = (6, 32, 8, 1)$ and $(9, 128, 32, 1)$ consistently improve with increasing $V$.

\begin{table}[tbp]
\caption{The $D1$ performance, the number of voxels, $|\mathcal{V}|$, and the number of bits to represent $\mathcal{O}_B$ w.r.t. different $B$.}
\begin{center}
\begin{tabular}{c|c|c|c}
\hline

\cline{1-4} 
\textbf{Block resolution $B$} & \textbf{1}& \textbf{2}& \textbf{3} \\
\hline

$D1$ (dB) & 45.52  & 45.82 & 46.18 \\
$|\mathcal{V}|$ & $1.6\times 10^6$ & $7.5 \times 10^5$  & $4.0\times 10^5$ \\
$3BN_B$ & 6 & 162 & 891\\

\hline
\end{tabular}
\label{tab:blk_res}
\end{center}

\end{table}

Then we investigate the effect of $B$ for $V = 7$. As analyzed in Section \ref{sec:IV}, adopting a larger $B$ value reduces the number of voxels within $\mathcal{V}$ and thus improves the reconstruction quality as well as the processing efficiency at the receiver. This is verified in the simulations where three neural networks with parameters $(L, d_1, d_2, N_r) = (6, 32, 8, 1)$ are trained with different block resolutions, $B = \{1, 2, 3\}$ over the same point cloud in Fig. \ref{fig:num_voxels}. The $D1$ performance, the number of voxels, $|\mathcal{V}|$, and the digital overhead to transmit the block occupancy information, $\mathcal{O}_B$, w.r.t. different $B$ are provided in Table \ref{tab:blk_res}.

As can be seen, with a large $B$, the reconstruction performance, $D1$, improves while the number of voxels, $|\mathcal{V}|$, reduces leading to a more efficient reconstruction process.
However,  the number of occupied blocks, $N_B$, grows rapidly with $B$ leading to excessive digital overhead to transmit the block occupancy information, $\mathcal{O}_B$. In particular, suppose there are $N_B \le 2^{3B-1}$ number of occupied blocks\footnote{When $N_B > 2^{3B-1}$, we will transmit the indices of the empty blocks instead.}, then, we need to transmit $3BN_B$ bits to the receiver over the wireless channel, where $3B$ is the number of bits to represent the index of each block.
To guarantee ultra reliable transmission over a wide range of channel SNRs would require introducing significant amount of redundancy, which in turn, increases the bandwidth cost of metadata.
Thus, we adopt $B = 2$ to strike a balance between the reconstruction performance and the channel bandwidth.

\begin{figure*}[t]
     \centering
      \begin{subfigure}{2\columnwidth}
         \centering
         \includegraphics[width=\columnwidth]{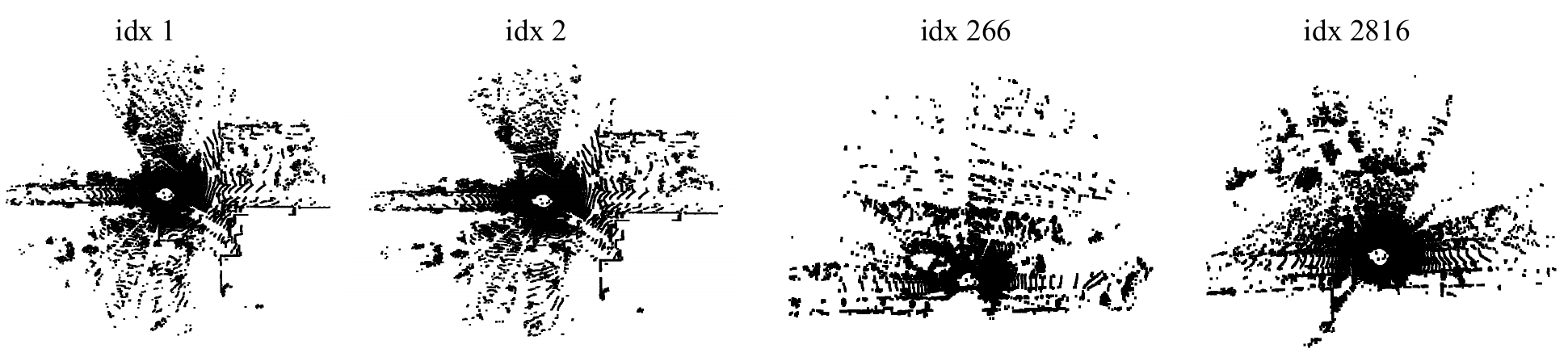}
         \caption{}
     \end{subfigure}    

     \begin{subfigure}{{0.6\columnwidth}}
         \centering
         \includegraphics[width=\columnwidth]{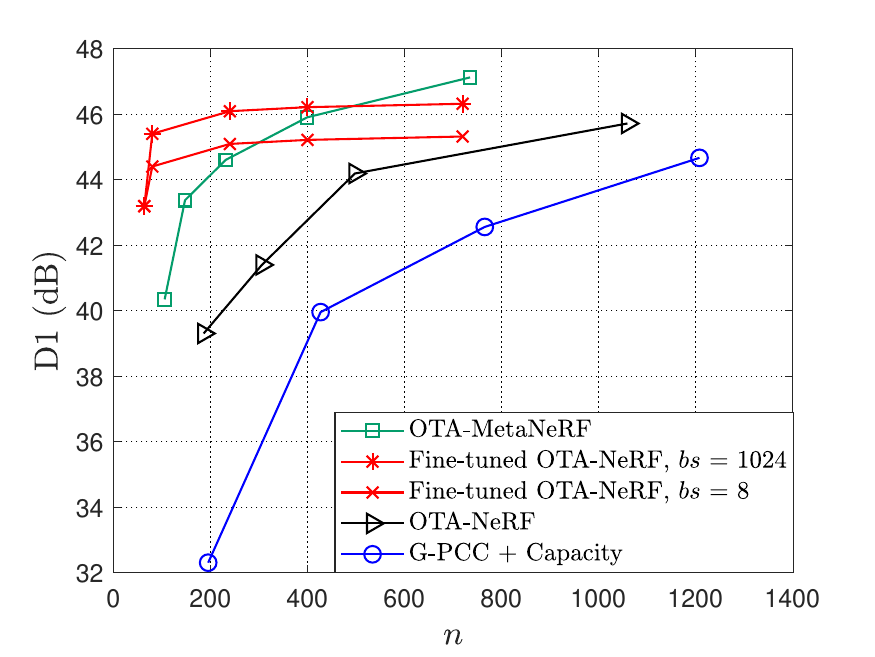}
         \caption{idx 2}
     \end{subfigure}
     \begin{subfigure}{{0.6\columnwidth}}
         \centering
         \includegraphics[width=\columnwidth]{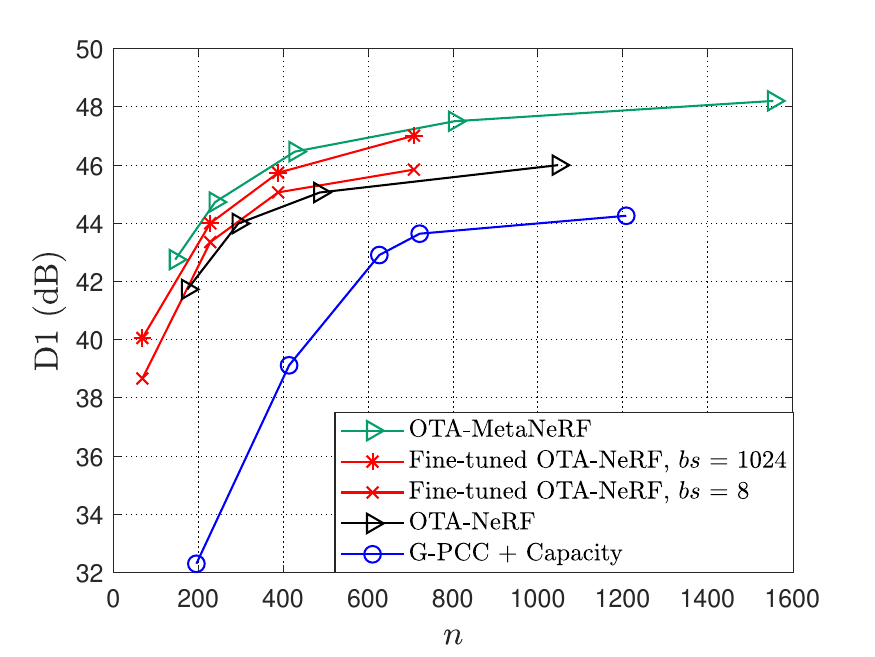}
         \caption{idx 266}
     \end{subfigure}
     \begin{subfigure}{0.6\columnwidth}
         \centering
         \includegraphics[width=\columnwidth]{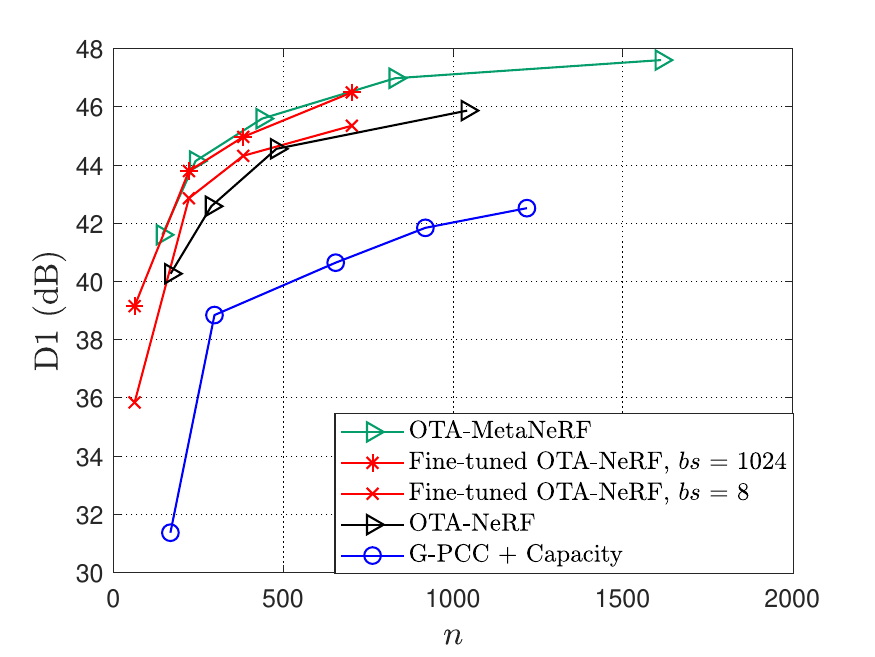}
         \caption{idx 2816}
     \end{subfigure}

  \caption{{Performance comparison of the proposed OTA-MetaNeRF,  OTA-NeRF with/without fine-tuning and the G-PCC baseline delivered at AWGN channel capacity. We set $k' = \{4, 8, 16, 32, 64\}$ to achieve different number of complex channel uses for the OTA-MetaNeRF scheme. For the fine-tuned OTA-NeRF scheme, the base model with parameters, $(L, d_1, d_2, N_r) = (6, 32, 8, 1)$, is trained on the point cloud with index 1 and evaluated on the point clouds with indices 2, 266 and 2816. Different fine-tuning parameters, $K = \{0, 8, 16, 32\}$, are considered.  We also provide visualization of the point clouds with different indices}. }
  \label{fig:ft_sim}
\end{figure*}

\begin{table}[tbp]
\caption{$D1$ performances of the proposed OTA-NeRF framework and the G-PCC baseline as a function of the channel SNR over a Rayleigh fading channel.}
\begin{center}
\begin{tabular}{c|ccccc}
\hline

\cline{1-4} 
\textbf{$\mathrm{SNR}$} (dB) & \textbf{5}& \textbf{7.5}& \textbf{10} & \textbf{12.5} & \textbf{15}\\
\hline

\textit{OTA-NeRF} (dB) & 41.76 & 44.95 & 45.03 & 45.79 & 46.01\\
\textit{G-PCC baseline} (dB) & 36.92 & 38.01 & 38.93 & 40.05 & 40.72\\

\hline
\end{tabular}
\label{tab:fd_D1}
\end{center}

\end{table}

\begin{table*}[ht]
\caption{{Run-time complexities of the proposed fine-tuned OTA-NeRF, OTA-MetaNeRF schemes and the G-PCC baseline.}}
\centering
{
\begin{tabular}{|c|ccc|cc|cc|}
\hline
\multirow{2}{*}{Schemes} & \multicolumn{3}{c|}{Fine-tuned OTA-NeRF} & \multicolumn{2}{c|}{OTA-MetaNeRF} & \multicolumn{2}{c|}{G-PCC} \\
\cline{2-8}
& encode ($bs = 1024$) & \multicolumn{1}{|c|}{encode ($bs = 8$)} & \multicolumn{1}{|c|}{decode} & encode & \multicolumn{1}{|c|}{decode} & encode & \multicolumn{1}{|c|}{decode}\\
\hline

Time (s) & 98.81 & \multicolumn{1}{|c|}{1.41} & \multicolumn{1}{|c|}{0.038} & 0.501 & \multicolumn{1}{|c|}{{0.046}} & 1.86 & \multicolumn{1}{|c|}{0.061} \\

\hline
\end{tabular}
}
\label{tab:ota_complex}
\vspace{0.2cm}

\end{table*}

\subsection{{Comparison of OTA schemes with Benchmarks}}
{We first compare the proposed OTA schemes, i.e., the OTA-NeRF and the OTA-MetaNeRF schemes with both conventional and learning-based point cloud transmission baselines over the AWGN channel, where the channel SNR is set to $\mathrm{SNR} = 10$ dB. For both OTA schemes, the reconstruction performances are evaluated over 10 arbitrarily chosen point clouds. 
The meta data containing the block occupancy information, $\mathcal{O}_B$, and the optimal number of reconstructed points, $N_p$, are transmitted using 1/2-rate Polar code with QPSK modulation for reliable transmission.
For the OTA-NeRF scheme, a neural network is overfitted to each point cloud, and { different numbers of channel uses are obtained by considering parameters $(L, d_1, d_2, N_r) = (2, 16, 4, 1), (4, 16, 4, 2), (6, 32, 8, 1), (6, 64, 16, 1)$, $(6, 64, 16, 2)$ and $(9, 128, 32, 1)$. }
For the OTA-MetaNeRF scheme, we set the number of SIREN layers, its hidden dimension, the parameter of the sinusoidal activation and the number of frequencies in the positional embedding layer to $(J, d, \omega_0, L) = (10, 256, 50, 6)$, respectively. The learning rate for the latent vector, $\bm{\phi}$, and the meta network parameters, $\bm{\Theta}$, are set to $10^{-2}$ and $3\times 10^{-6}$, respectively. $T = 3$ optimization steps are performed to update the latent vectors in both training and evaluation phases. Different numbers of complex channel uses are achieved by employing $k' = \{4, 8, 16, 32, 64, 128, 256\}$.}
We first consider two baselines using conventional point cloud compression algorithms, namely, Draco \cite{GoogleDraco} and G-PCC \cite{GPCC}, whose bit outputs are protected by rate-1/2 Polar code followed by 16QAM modulation. Two deep learning based baselines are implemented, the first one is the SEPT in Section \ref{sec:III} and the other is the `depoco' proposed in \cite{wiesmann2021deep}. 

As shown in Fig. \ref{fig:final_simu}(a), the proposed OTA-NeRF framework outperforms all baselines for $n \le 8\times 10^3$. However, its reconstruction performance improves slowly when $n > 4\times 10^3$. {This can be understood from two perspectives. Firstly, there exists a reconstruction upper bound, i.e., $D1 \approx 50$ dB, as both the OTA schemes adopt voxelization with $V = 6$, which leads to a loss of details. Secondly, by the neural scaling laws \cite{neural_scale}, the MLP layers of the OTA-NeRF model can only achieve a smaller test error with a larger number of parameters. The OTA-MetaNeRF scheme achieves a superior reconstruction performance over the OTA-NeRF scheme, approaching the upper bound with latent dimension $k^\prime = 256$.} It is also observed that though SEPT can achieve satisfactory reconstruction performance on the downsampled ShapeNet dataset with small number of points, it falls short when dealing with larger point clouds. This aligns with the intuition that the geometric information is lost due to the max-pooling operation leading to less satisfactory reconstruction performance. The depoco baseline transmits its downsampled coordinates using digital coded modulation, which not only achieves a lower $D1$ value on average, but also suffers from cliff and leveling effects.

We then show that the superiority of {both OTA-NeRF and OTA-MetaNeRF schemes} extends to different SNR values. We compare the proposed OTA-NeRF and OTA-MetaNeRF models with the G-PCC baseline due to its superior reconstruction performance in Fig. \ref{fig:final_simu}(a). {In particular, for the OTA-NeRF scheme, we set $(L, d_1, d_2, N_r) = (6, 32, 8, 1)$ leading to $1993$ learnable parameters.  We assume the meta data is transmitted at a rate equal to the AWGN channel capacity at $\mathrm{SNR} = 0$ dB, and the total number of complex channel uses is $n = 1147$.  For the OTA-MetaNeRF scheme, on the other hand, we set $k' = 32$ leading to $n = 774$, which is $70\%$ of the bandwidth used by the OTA-NeRF scheme.
We also provide the $D1$ and $D2$ performances of the OTA-NeRF and OTA-MetaNeRF models trained at $5$ dB and evaluated under different SNR values.}
For the G-PCC baseline, we consider transmitting its compression output using different coding rates. In particular, we consdier a rate-1/2 Polar code with 16QAM,  rate-3/4 Polar code with 16QAM, and a capacity-achieving coding scheme. 
The relative performances of the aforementioned schemes are shown in Fig. \ref{fig:final_simu} (b) \& (c).

As can be seen from these figures, both $D1$ and $D2$ performances achieved by the proposed OTA-NeRF and OTA-MetaNeRF models trained and tested at the same SNR outperform that of the G-PCC baselines adopting different coded modulation schemes, including the capacity-achieving baseline. 
{It is worth mentioning that the OTA-MetaNeRF scheme slightly outperforms OTA-NeRF despite using 400 fewer complex channel symbols.
Moreover, both the OTA-NeRF and OTA-MetaNeRF models trained at fixed $\mathrm{SNR} = 5$ dB yield satisfactory reconstruction performance when evaluated at other SNR values, which illustrates that the OTA schemes avoid the cliff and leveling effects.} The G-PCC baseline with digital coded modulation, on the other hand, completely fails to reconstruct the original point cloud when the SNR drops below a certain threshold, and its performance saturates beyond this threshold.

We then evaluate the reconstruction performance of the OTA-NeRF scheme over Rayleigh fading channels. In this simulation, we set the neural network parameters for the OTA-NeRF model as ${(L, d_1, d_2, N_r)} = (6, 32, 8, 1)$. In the G-PCC baseline, the compressed bits are delivered at the ergodic capacity  of the fading channel. Their reconstruction performances are shown in Table \ref{tab:fd_D1}. As can be seen from the table, the proposed OTA-NeRF scheme significantly outperforms the G-PCC baseline, even though the latter is assumed to communicate at the ergodic capacity.

\subsection{{Real-Time OTA Schemes}}
{Finally, we perform numerical experiments to illustrate the effectiveness of the proposed fine-tuned OTA-NeRF and the OTA-MetaNeRF schemes, as well as their capabilities for real-time point cloud communications. The simulations are carried out on the same machine equipped with RTX 3080 GPU as in the SEPT model}.  Due to the page limit, we show the effectiveness of the proposed OTA schemes using the point clouds with indices $2, 266$ and $2816$ over the AWGN channel with $\mathrm{SNR} = 10$ dB.

For the fine-tuned OTA-NeRF scheme, we first train a base model on the first point cloud whose parameters are set to $(L, d_1, d_2, N_r) = (6, 32, 8, 1)$. The base model is assumed to be available to the receiver in advance and only the updated parameters need to be transmitted over the wireless channel for different point cloud inputs as illustrated in Section \ref{sec:finetune}.
The fine-tuning parameter is set to $K = \{0, 8, 16, 32\}$ leading to different numbers of complex channel uses. {The model is trained for 10 epochs using $bs$ number of batches for each epoch with batch size equals to $16384$. A larger $bs$ leads to improved reconstruction performance yet requires higher computational overhead.} 
{The OTA-MetaNeRF scheme achieves different number of channel uses by setting the dimension of the latent vectors as  $k' \in \{4, 8, 16, 32, 64\}$.}
For the original OTA-NeRF models illustrated in Section \ref{sec:OTA-NeRF}, we consider four sets of parameters, $(L, d_1, d_2, N_r) = (2, 8, 4, 1), (2, 16, 4, 1), (6, 16, 4, 1)$ and $(6, 32, 8, 1)$.  The $D1$ performances as well as the run-time complexities obtained by the OTA-MetaNeRF, fine-tuned/original OTA-NeRF models, and the G-PCC baseline delivered at AWGN channel capacity are shown in Fig. \ref{fig:ft_sim} and Table \ref{tab:ota_complex}, respectively. 

As can be seen, for all the aforementioned point clouds, the $D1$ performance of the fine-tuned OTA-NeRF models outperform the original ones trained from scratch. {Moreover, the fine-tuned OTA-NeRF models with $bs = \{8, 1024\}$ outperform the G-PCC benchmark by $n \approx 650, 400, 750$ complex channel uses for the three point clouds when we set the desired reconstruction performance to $D1 = 42$ dB. It is worth mentioning that adopting $bs = 8$ is enough as it is only slightly outperformed by $bs = 1024$ yet yields much lower encoding latency. 
As shown in Table \ref{tab:ota_complex}, both the encoding and decoding overheads of the $bs = 8$ case are less than those of the benchmark. However, its $bs = 1024$ counterpart has the same decoding overhead yet encodes the point cloud much slower taking $\sim 100s$ which is prohibitive for real-time applications.} We can also observe from the figure that the gain obtained by the fine-tuning scheme over the original one varies from one point cloud to the next. This can be understood from the visualizations in Fig. \ref{fig:ft_sim}: since the point cloud with index 2 is nearly identical to the one with index 1 for which the base model is trained, the reconstruction quality obtained even with $K = 0$ is good enough, and not much is gained by increasing $K$ in this case. On the other hand, the point clouds with indices $266$ and $2816$ appear to be quite different, and hence, they require higher $K$ values, but they are still more efficient compared to training a neural network from scratch.

{Finally, we focus on the OTA-MetaNeRF scheme. As shown in Fig. \ref{fig:ft_sim}, it achieves the best reconstruction performance for the point clouds with indices $266$ and $2816$, and is only outperformed by the fine-tuned OTA-NeRF scheme when $n<400$ for the point cloud with index 2. Moreover, it achieves the lowest encoding  and the second lowest decoding delays as in Table \ref{tab:ota_complex}. These manifest the effectiveness of the proposed OTA-MetaNeRF scheme, particularly its generalization capability to different point clouds. }

\section{Conclusion}
{In this paper, we have studied the timely and challenging problem of wireless point cloud transmission, and proposed three novel frameworks, namely, SEPT, OTA-NeRF and OTA-MetaNeRF.} For relatively small point clouds, the SEPT framework is shown to enable efficient, robust {and SNR-adaptive} delivery. {For larger point clouds, we introduced the OTA-NeRF and OTA-MetaNeRF frameworks, which encode the point cloud data into the weights of a neural network or the latent vectors, respectively.} A fine-tuned OTA-NeRF algorithm is proposed where only a fraction of the neural network weights are re-trained and transmitted, which significantly reduces the bandwidth cost, particularly when there is correlation among the transmitted point clouds. {The OTA-MetaNeRF scheme employs a meta-learning approach, and improves the reconstruction performance, the encoding efficiency and the generality w.r.t. the OTA-NeRF scheme.} Extensive numerical experiments are carried out to verify that all the proposed frameworks achieve superior or comparable reconstruction performances w.r.t existing point cloud transmission schemes while mitigating the cliff and leveling effects. {Simulations are carried out to confirm the capability of the proposed schemes for real-time point cloud transmission}.

\appendices


\bibliographystyle{IEEEtran}
\bibliography{References}

\begin{thebibliography}{10}
\providecommand{\url}[1]{#1}
\csname url@samestyle\endcsname
\providecommand{\newblock}{\relax}
\providecommand{\bibinfo}[2]{#2}
\providecommand{\BIBentrySTDinterwordspacing}{\spaceskip=0pt\relax}
\providecommand{\BIBentryALTinterwordstretchfactor}{4}
\providecommand{\BIBentryALTinterwordspacing}{\spaceskip=\fontdimen2\font plus
\BIBentryALTinterwordstretchfactor\fontdimen3\font minus \fontdimen4\font\relax}
\providecommand{\BIBforeignlanguage}[2]{{%
\expandafter\ifx\csname l@#1\endcsname\relax
\typeout{** WARNING: IEEEtran.bst: No hyphenation pattern has been}%
\typeout{** loaded for the language `#1'. Using the pattern for}%
\typeout{** the default language instead.}%
\else
\language=\csname l@#1\endcsname
\fi
#2}}
\providecommand{\BIBdecl}{\relax}
\BIBdecl

\bibitem{sept}
C.~Bian, Y.~Shao, and D.~G{\"u}nd{\"u}z, ``Wireless point cloud transmission,'' in \emph{IEEE Int'l Wrksp. Signal Proc. Advances in Wireless Comms. (SPAWC)}, 2024.

\bibitem{Rusu_ICRA2011_PCL}
R.~B. Rusu and S.~Cousins, ``{3\textsc{D} is here: Point Cloud Library},'' in \emph{{IEEE Int. Conf. Robot. Autom.}}, 2011.

\bibitem{qi2017pointnet}
C.~R. Qi, H.~Su, K.~Mo, and L.~J. Guibas, ``Pointnet: Deep learning on point sets for 3\textsc{D} classification and segmentation,'' in \emph{CVPR}, 2017, pp. 652--660.

\bibitem{ptv1}
H.~Zhao, L.~Jiang, J.~Jia, P.~H. Torr, and V.~Koltun, ``Point transformer,'' in \emph{ICCV}, 2021, pp. 16\,259--16\,268.

\bibitem{thomas2019kpconv}
H.~Thomas, C.~R. Qi, J.-E. Deschaud, B.~Marcotegui, F.~Goulette, and L.~J. Guibas, ``Kpconv: Flexible and deformable convolution for point clouds,'' in \emph{CVPR}, 2019, pp. 6411--6420.

\bibitem{guo2021pct}
M.-H. Guo, J.-X. Cai, Z.-N. Liu, T.-J. Mu, R.~R. Martin, and S.-M. Hu, ``\textsc{Pct}: Point cloud transformer,'' \emph{Computational Visual Media}, vol.~7, pp. 187--199, 2021.

\bibitem{fpt}
C.~Park, Y.~Jeong, M.~Cho, and J.~Park, ``Fast point transformer,'' in \emph{CVPR}, 2022, pp. 16\,949--16\,958.

\bibitem{OctSqueeze}
L.~Huang, S.~Wang, K.~Wong, J.~Liu, and R.~Urtasun, ``Oct\textsc{S}queeze: Octree-structured entropy model for \textsc{L}i\textsc{DAR} compression,'' in \emph{CVPR}, June 2020.

\bibitem{pcgcv2}
J.~Wang, D.~Ding, Z.~Li, and Z.~Ma, ``Multiscale point cloud geometry compression,'' in \emph{IEEE Data Compression Conf. (DCC)}, 2021, pp. 73--82.

\bibitem{wiesmann2021deep}
L.~Wiesmann, A.~Milioto, X.~Chen, C.~Stachniss, and J.~Behley, ``Deep compression for dense point cloud maps,'' \emph{IEEE Robot. Autom. Lett.}, vol.~6, no.~2, pp. 2060--2067, 2021.

\bibitem{pct_pcc}
J.~Zhang, G.~Liu, D.~Ding, and Z.~Ma, ``Transformer and upsampling-based point cloud compression,'' in \emph{Proc. 1st Int. Workshop Adv. Point Cloud Compr., Process. Anal.}, 2022, pp. 33--39.

\bibitem{ruan2024pointcloudcompressionimplicit}
H.~Ruan, Y.~Shao, Q.~Yang, L.~Zhao, and D.~Niyato, ``Point cloud compression with implicit neural representations: A unified framework,'' \emph{arXiv:2405.11493}, 2024.

\bibitem{nvfpcc}
Y.~Hu and Y.~Wang, ``Learning neural volumetric field for point cloud geometry compression,'' in \emph{Picture Coding Symposium (PCS)}, 2022, pp. 127--131.

\bibitem{NeRF}
B.~Mildenhall, P.~P. Srinivasan, M.~Tancik, J.~T. Barron, R.~Ramamoorthi, and R.~Ng, ``\textsc{NeRF}: representing scenes as neural radiance fields for view synthesis,'' \emph{Commun. ACM}, vol.~65, no.~1, p. 99–106, Dec 2021.

\bibitem{coin}
E.~Dupont, A.~Goliński, M.~Alizadeh, Y.~W. Teh, and A.~Doucet, ``Coin: Compression with implicit neural representations,'' \emph{arXiv:2103.03123}, 2021.

\bibitem{deepcabac}
S.~Wiedemann, H.~Kirchhoffer, S.~Matlage, P.~Haase, A.~Marban, T.~Marinč, D.~Neumann, T.~Nguyen, H.~Schwarz, T.~Wiegand, D.~Marpe, and W.~Samek, ``Deepcabac: A universal compression algorithm for deep neural networks,'' \emph{IEEE J. Sel. Topics Signal Process.}, vol.~14, no.~4, pp. 700--714, 2020.

\bibitem{GPCC}
D.~Graziosi, O.~Nakagami, S.~Kuma, A.~Zaghetto, T.~Suzuki, and A.~Tabatabai, ``An overview of ongoing point cloud compression standardization activities: Video-based (\textsc{V}-\textsc{PCC}) and geometry-based (\textsc{G}-\textsc{PCC}),'' \emph{APSIPA Trans. Signal Info. Process.}, vol.~9, pp. 1--13, 2020.

\bibitem{coinpp}
E.~Dupont, H.~Loya, M.~Alizadeh, A.~Goliński, Y.~W. Teh, and A.~Doucet, ``Coin++: Neural compression across modalities,'' \emph{arXiv:2201.12904}, 2022.

\bibitem{gündüz2024jointsourcechannelcodingfundamentals}
D.~G{\"u}nd{\"u}z, M.~A. Wigger, T.-Y. Tung, P.~Zhang, and Y.~Xiao, ``Joint source–channel coding: Fundamentals and recent progress in practical designs,'' \emph{Proc. IEEE}, pp. 1--32, 2024.

\bibitem{JSCC2019}
E.~Bourtsoulatze, D.~B. Kurka, and D.~G{\"u}nd{\"u}z, ``Deep joint source-channel coding for wireless image transmission,'' \emph{IEEE Trans. Cogn. Commun. Netw.}, vol.~5, no.~3, pp. 567--579, 2019.

\bibitem{Deniz2022}
D.~G{\"u}nd{\"u}z, Z.~Qin, I.~E. Aguerri, H.~S. Dhillon, Z.~Yang, A.~Yener, K.~K. Wong, and C.-B. Chae, ``Beyond transmitting bits: Context, semantics, and task-oriented communications,'' \emph{IEEE J. Sel. Area Commun.}, 2022.

\bibitem{jsccofdm}
M.~Yang, C.~Bian, and H.-S. Kim, ``\textsc{OFDM}-guided deep joint source channel coding for wireless multipath fading channels,'' \emph{IEEE Trans. Cogn. Commun. Netw.}, vol.~8, no.~2, pp. 584--599, 2022.

\bibitem{vit_mimo}
H.~Wu, Y.~Shao, C.~Bian, K.~Mikolajczyk, and D.~Gündüz, ``Deep joint source-channel coding for adaptive image transmission over \textsc{MIMO} channels,'' \emph{IEEE Trans. Wireless Commun.}, vol.~23, no.~10, pp. 15\,002--15\,017, 2024.

\bibitem{pf_relay_jscc}
C.~Bian, Y.~Shao, H.~Wu, E.~Ozfatura, and D.~G{\"u}nd{\"u}z, ``Process-and-forward: Deep joint source-channel coding over cooperative relay networks,'' \emph{IEEE J. Sel. Area Commun.}, 2025.

\bibitem{bupt}
J.~Dai, S.~Wang, K.~Tan, Z.~Si, X.~Qin, K.~Niu, and P.~Zhang, ``Nonlinear transform source-channel coding for semantic communications,'' \emph{IEEE J. Sel. Area Commun.}, vol.~40, no.~8, pp. 2300--2316, 2022.

\bibitem{deepwive}
T.-Y. Tung and D.~G{\"u}nd{\"u}z, ``Deep\textsc{W}i\textsc{V}e: Deep-learning-aided wireless video transmission,'' \emph{IEEE J. Sel. Areas Commun.}, vol.~40, no.~9, pp. 2570--2583, 2022.

\bibitem{deepjscc_speech}
M.~Bokaei, J.~Jensen, S.~Doclo, and J.~Østergaard, ``Low-latency deep analog speech transmission using joint source channel coding,'' \emph{IEEE J. Sel. Topics Signal Process.}, vol.~18, no.~8, pp. 1401--1413, 2024.

\bibitem{airnet}
M.~Jankowski, D.~Gündüz, and K.~Mikolajczyk, ``\textsc{AirNet}: Neural network transmission over the air,'' \emph{IEEE Trans. Wireless Commun.}, vol.~23, no.~9, pp. 12\,126--12\,139, 2024.

\bibitem{pointcloudvideo}
Y.~Huang, B.~Bai, Y.~Zhu, X.~Qiao, X.~Su, L.~Yang, and P.~Zhang, ``\textsc{ISCom}: Interest-aware semantic communication scheme for point cloud video streaming on metaverse \textsc{XR} devices,'' \emph{IEEE J. Sel. Areas Commun.}, vol.~42, no.~4, pp. 1003--1021, 2024.

\bibitem{HoloCast+}
T.~Fujihashi, T.~Koike-Akino, T.~Watanabe, and P.~V. Orlik, ``Holocast+: Hybrid digital-analog transmission for graceful point cloud delivery with graph \textsc{F}ourier transform,'' \emph{IEEE Trans. Multimedia}, vol.~24, pp. 2179--2191, 2022.

\bibitem{gnn_pct}
T.~Fujihashi, T.~Koike-Akino, S.~Chen, and T.~Watanabe, ``Wireless 3d point cloud delivery using deep graph neural networks,'' in \emph{ICC 2021 - IEEE Int. Conf. Commun.}, 2021, pp. 1--6.

\bibitem{jap_inr}
T.~Fujihashi, S.~Kato, and T.~Koike-Akino, ``Implicit neural representation for low-overhead graph-based holographic-type communications,'' in \emph{Proc. IEEE Int. Conf. Acoust. Speech Signal Process. (ICASSP)}, 2024, pp. 2825--2829.

\bibitem{ext_sept}
S.~Xie, Q.~Yang, Y.~Sun, T.~Han, Z.~Yang, and Z.~Shi, ``Semantic communication for efficient point cloud transmission,'' \emph{arXiv:2409.03319}, 2024.

\bibitem{xu2021wireless}
J.~Xu, B.~Ai, W.~Chen, A.~Yang, P.~Sun, and M.~Rodrigues, ``Wireless image transmission using deep source channel coding with attention modules,'' \emph{IEEE Trans. Circuits Syst. Video Technol.}, 2021.

\bibitem{D1D2}
``Common test conditions for point cloud compression,'' \emph{ISO/IEC JTC1/SC29/WG11 MPEG output document N19084}, Feb. 2020.

\bibitem{chang2015shapenet}
A.~X. Chang, T.~Funkhouser, L.~Guibas, P.~Hanrahan, Q.~Huang, Z.~Li, S.~Savarese, M.~Savva, S.~Song, H.~Su \emph{et~al.}, ``Shapenet: An information-rich 3\textsc{D} model repository,'' \emph{arXiv preprint arXiv:1512.03012}, 2015.

\bibitem{finite_length_capacity}
Y.~Polyanskiy, H.~V. Poor, and S.~Verdu, ``Channel coding rate in the finite blocklength regime,'' \emph{IEEE Trans. Info. Theory}, vol.~56, no.~5, pp. 2307--2359, 2010.

\bibitem{GoogleDraco}
Google, ``Draco 3d data compression,'' \url{https://github.com/google/draco}.

\bibitem{neural_scale}
J.~Kaplan, S.~McCandlish, T.~Henighan, T.~B. Brown, B.~Chess, R.~Child, S.~Gray, A.~Radford, J.~Wu, and D.~Amodei, ``Scaling laws for neural language models,'' \emph{arXiv preprint arXiv:2001.08361}, 2020.

\end{thebibliography}

\begin{IEEEbiography}[{\includegraphics[width=1.1in,height=1.3in,clip,keepaspectratio]{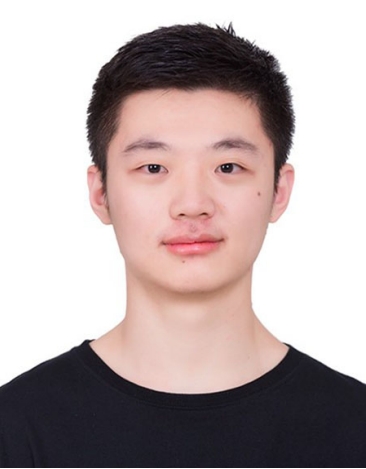}}]{Chenghong Bian} received the B.S. degree from the Physics Department, Tsinghua University, in 2020, and the M.S. degree from the EECS Department, University of Michigan, in 2022. He is currently pursuing the Ph.D. degree with the Department of Electrical and Electronic Engineering, Imperial College London. His research interests include wireless communications, machine learning and sensing.  He received the Best Paper Award at IEEE International Conference on Communications (ICC) 2023.
\end{IEEEbiography}

\begin{IEEEbiography}[{\includegraphics[width=1.1in,height=1.3in,clip,keepaspectratio]{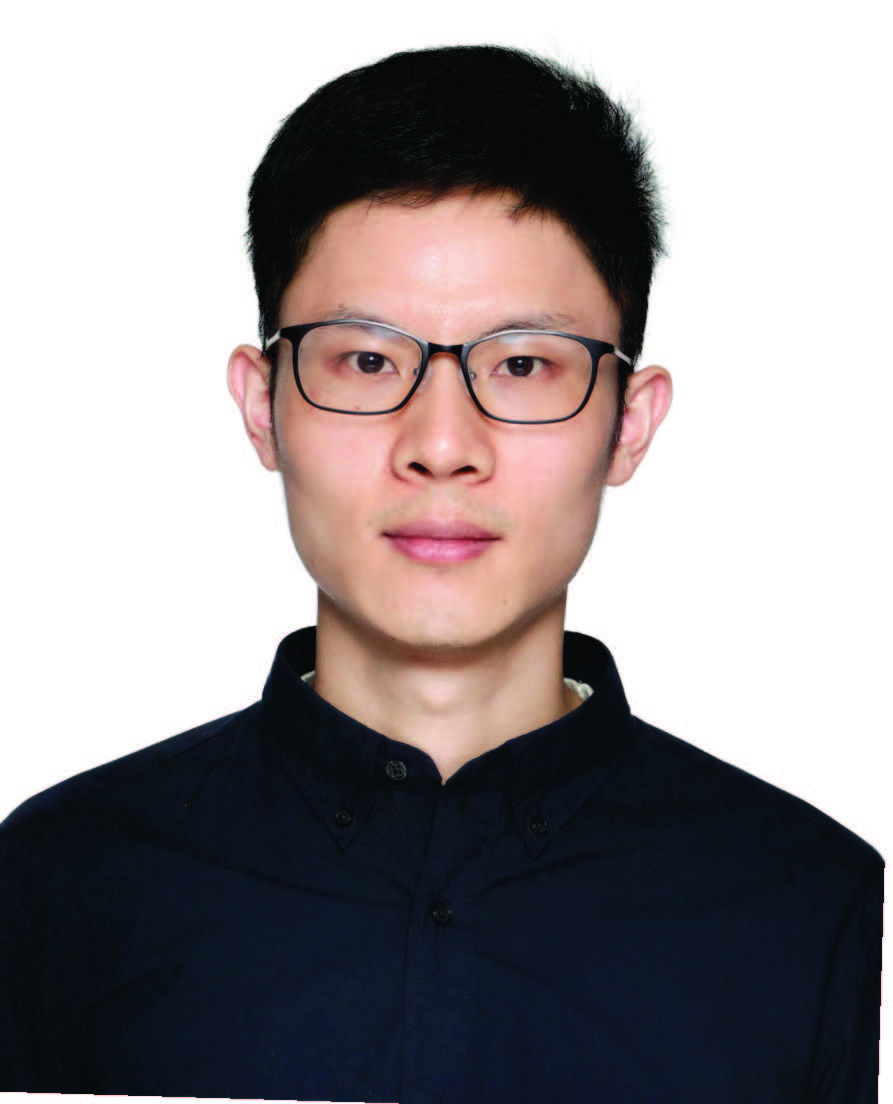}}]{Yulin Shao} (Member, IEEE) is an Assistant Professor with the Department of Electrical and Electronic Engineering, The University of Hong Kong (HKU). He received the B.S. and M.S. degrees in Communications and Information Engineering (Hons.) from Xidian University, China, in 2013 and 2016, respectively, and the Ph.D. degree in Information Engineering from The Chinese University of Hong Kong (CUHK) in 2020. He was a Research Assistant with the Institute of Network Coding (INC), a Visiting Scholar with the Research Laboratory of Electronics at Massachusetts Institute of Technology (MIT), a Research Associate with the Department of Electrical and Electronic Engineering at Imperial College London (ICL), and a Lecturer in Information Processing with the University of Exeter. He was a Guest Lecturer at 5G Academy Italy and IEEE Information Theory Society Bangalore Chapter.

Dr. Shao's research interests include coding and modulation, machine learning, and stochastic control. He is a Series Editor of IEEE Communications Magazine in the area of Artificial Intelligence and Data Science for Communications, an Editor of IEEE Transactions on Communications in the area of Machine Learning and Communications, and an Editor of IEEE Communications Letters. He received the Best Paper Awards at IEEE International Conference on Communications (ICC) 2023, and IEEE Wireless Communications and Networking Conference (WCNC) 2024.
\end{IEEEbiography}

\begin{IEEEbiography}[{\includegraphics[width=1.1in,height=1.3in,clip,keepaspectratio]{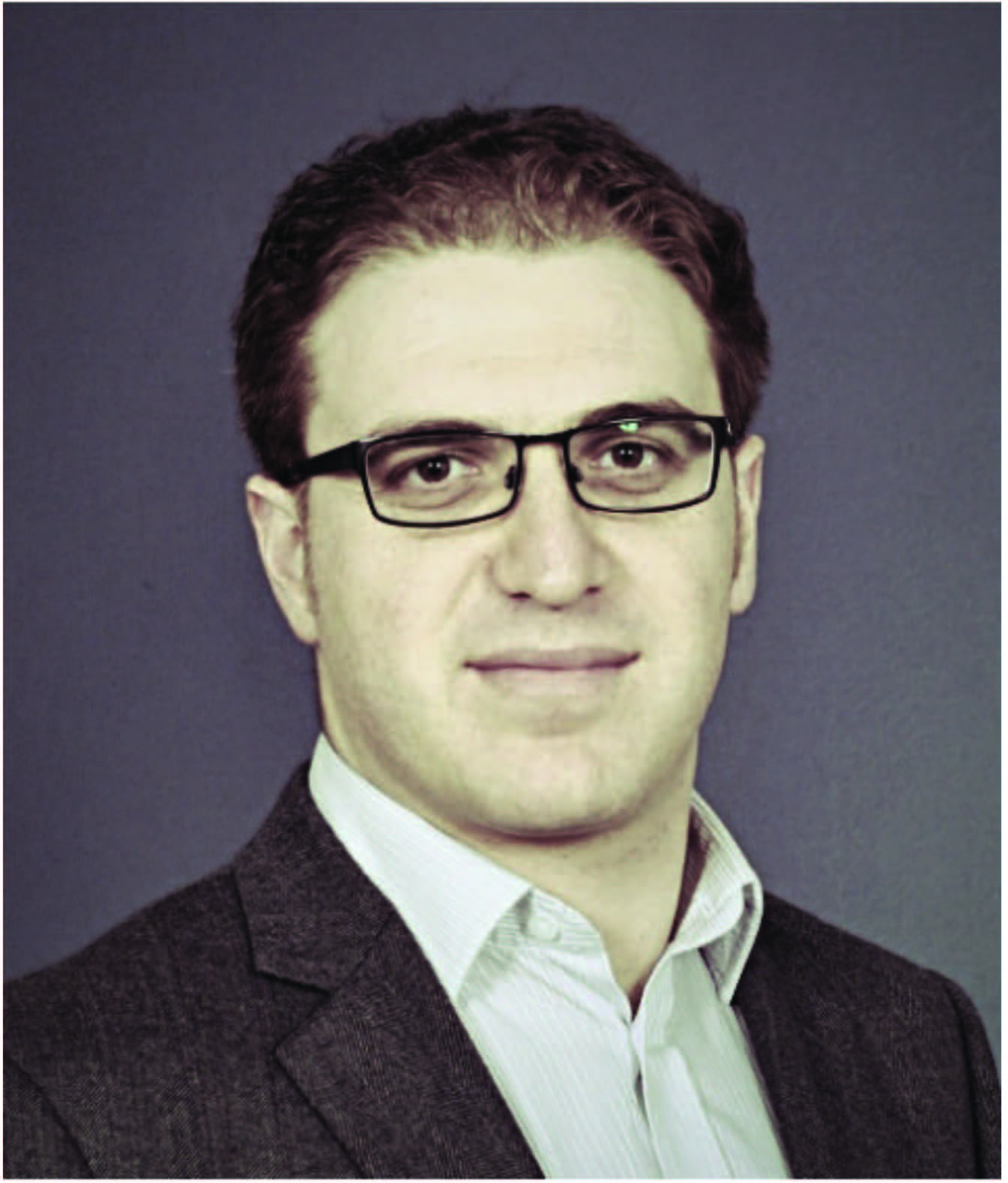}}]{Deniz~G\"und\"uz} (Fellow, IEEE) received the B.S. degree in electrical and electronics engineering from METU, Turkey in 2002, and the M.S. and Ph.D. degrees in electrical engineering from NYU Tandon School of Engineering (formerly Polytechnic University) in 2004 and 2007, respectively. Currently, he is a Professor of Information Processing in the Electrical and Electronic Engineering Department at Imperial College London, UK, where he also serves as the deputy head of the Intelligent Systems and Networks Group. In the past, he held various positions at the University of Modena and Reggio Emilia (part-time faculty member, 2019-22), University of Padova (visiting professor, 2018, 2020), Centre Tecnologic de Telecomunicacions de Catalunya (CTTC) (research associate, 2009-12), Princeton University (postdoctoral researcher, 2007-09, visiting researcher, 2009-11) and Stanford University (research assistant professor, 2007-09). His research interests lie in the areas of communications and information theory, machine learning, and privacy.

Deniz~G\"und\"uz is a Fellow of the IEEE. He is an elected member of the IEEE Signal Processing Society Signal Processing for Communications and Networking (SPCOM) and Machine Learning for Signal Processing (MLSP) Technical Committees. He serves as an Area Editor for the IEEE Transactions on Information Theory and IEEE Transactions on Communications. He is the recipient of the IEEE Communications Society Communication Theory Technical Committee (CTTC) Early Achievement Award in 2017, Starting (2016) and Consolidator (2022) and Proof-of-Concept (2023) Grants of the European Research Council (ERC), and has co-authored several award-winning papers, including the IEEE Communications Society - Young Author Best Paper Award (2022), and IEEE International Conference on Communications Best Paper Award (2023). He received the Imperial College London - President’s Award for Excellence in Research Supervision in 2023.
\end{IEEEbiography}

\end{document}